\documentclass[a4paper, 12pt]{article}

\usepackage[T1]{fontenc}
\usepackage[utf8]{inputenc}

\usepackage{times}
\usepackage{amsfonts, amssymb, amsthm, graphicx, mathtools, epstopdf}
\usepackage{accents}

\usepackage[colorlinks, citecolor=blue, urlcolor=blue]{hyperref}

\usepackage{graphicx, caption, subcaption}
\usepackage{microtype}
\usepackage{setspace}
\usepackage[margin=1in]{geometry}

\newtheorem{assumption}{Assumption}
\newtheorem{example}{Example}

\newcommand{\obf}{{\bf 1}}

\newcommand{\st}{{\textrm{such that }}}
\newcommand{\Var}{{\operatorname{Var\,}}}

\newcommand{\Wb}{{W_{\operatorname{basic}}}}
\newcommand{\Ws}{{W_{\operatorname{spill}}}}

\newcommand{\FC}{{\textrm{fc}}}
\newcommand{\FT}{{\textrm{ft}}}

\title{Estimation of Monotone Treatment Effects in Network Experiments}
\author{David Choi}
\begin{document}
\maketitle

\begin{abstract}
Randomized experiments on social networks pose statistical challenges, due to the possibility of interference between units. We propose  new methods for estimating attributable treatment effects in such settings. The methods do not require partial interference, but instead require an identifying assumption that is similar to requiring nonnegative treatment effects. Network or spatial information can be used to customize the test statistic; in principle, this can increase power without making assumptions on the data generating process. 
\end{abstract}

\doublespacing

\section{Introduction} \label{sec: intro}

Spillover effects, social influence, and the sharing of information are widely believed to be important mechanisms for social and economic systems. To better understand them, researchers may collect network data on relationships between units. In some cases, the data may come from a randomized experiment; past examples include studies in viral marketing \cite{aral2011creating}, voting behavior \cite{bond201261, nickerson2008voting}, online sharing \cite{kramer17062014}, education \cite{sweet2013hierarchical}, and health \cite{miguel2004worms}. 

In such experiments, the outcomes tend to be social in nature, and the treatment of one individual may influence others. This phenomenon, known as interference, often complicates the analysis. For example, \cite{bond201261} describes an experiment that  was conducted using Facebook, a social network website. On the day of the 2010 US midterm Congressional elections, participants received a banner advertisement on Facebook which encouraged them to vote, with the option to self-report that they had voted by clicking on an ``I voted'' button. This advertisement was customized for each recipient, so that it displayed the total number of users who had already viewed the advertisement and clicked ``I voted''; for a random subset, the advertisement also displayed the profile pictures of up to six of the recipient's Facebook friends who had already self-reported. The self-reported voting rate for the treatment group (those receiving profile pictures) was 2.08\% higher than for the other participants, a difference large enough to reject a sharp null of zero effect. Since the content of the advertisement for each viewer depended on the actions of previous viewers, the presence of peer effects was ensured by the experiment design. Additionally, participants may have influenced each other through conversations caused by viewing the advertisement. Due to this interference, rigorous estimates of the effect size do not necessarily follow from rejection of the sharp null, as estimation methods that assume no interference may not be applicable.


We propose a new approach for these types of experiments, which is based on an identifying assumption that the treatment effect is monotone. This is slightly weaker than requiring the treatment to not have negative effects, either directly or indirectly, on the outcome of any unit. Aside from this assumption, the interference will be allowed to take arbitrary and unknown form. Specifically, we do not assume partial interference or a correctly specified model of social influence. 

The outline of the paper is as follows. Section \ref{sec: literature} surveys related works. The basic problem formulation is given in Section \ref{sec: notation}. Three methods for estimation are presented in Section \ref{sec: methods}. These methods are demonstrated using data and simulation examples in Section \ref{sec: simulation}. Section \ref{sec: discussion} discusses practical issues and future directions. Further technical details of the methods are presented in the appendices.



\section{Related Work} \label{sec: literature}

Early discussion of interference in the potential outcomes framework is attributed to \cite{rubin1990comment, halloran1995causal}. Current methods can be broadly divided between those which use a distribution-free rank statistic, and those which add identifying assumptions. 

Distribution-free rank statistics are considered in \cite{rosenbaum2007interference, luo2012inference}. In this approach, no assumptions are made on the interference, so that the estimates are highly robust. However, estimation is limited to rank-based quantities, i.e., on whether the treatment caused an overall shift in the ranks of the treated population when ordering the units by outcome. For non-rank quantities of interest, such as the average outcome under a counterfactual treatment, it appears that additional assumptions are required.

The most common identifying assumption is that the units form groups (such as households or villages) that do not interfere with each other; this is termed partial interference \cite{sobel2006randomized}. The paper \cite{hudgens2008toward} derives unbiased point estimates under partial interference, and variance bounds on the estimation error under a stronger condition termed stratified interference. Asymptotically normal estimates are given in \cite{liu2013large}, again assuming stratified interference, and finite sample error bounds are derived in \cite{tchetgen2012causal}. For settings where partial interference does not apply, more general exposure models have been investigated by \cite{toulis2013estimation, ugander2013graph, aronow2012estimating, ogburn2014vaccines, manski2013identification}, with rigorous results if one assumes knowledge of the network dynamics, such as who influences whom. As a result, they may not be suitable when the underlying social mechanisms are not well understood. The recent paper \cite{eckles2014design} also studies biased estimation of treatment effects under weaker assumptions than partial or fully modeled interference, which is similar in spirit to this present work.


\section{Setup and notation} \label{sec: notation}

Let $N$ denote the number of units in the experiment. Let treatments be assigned by sampling $L$ units without replacement, and let $X = (X_1,\ldots,X_N)$ encode the treatment assignment, where $X_i = 1$ if the $i$th unit was selected for treatment and $X_i=0$ otherwise. Let $Y =(Y_1,\ldots,Y_N)$ denote the observed outcomes, and let $\theta = (\theta_1,\ldots,\theta_N)$ denote the counterfactual outcomes under ``full control'', i.e., if none of the units had received treatment and $X_i=0$ for all $i$. 

As previously mentioned, we do not require an assumption of partial interference to hold. Instead, we require the following assumption on the treatment effect:
\begin{assumption}[Monotonicity] \label{as: monotone}
$\theta_i \leq Y_i$, for all $i=1,\ldots,N$.
\end{assumption}
This assumption might not be appropriate for some applications; for example, police interventions might displace crime, so that crime rates would decrease in some areas but increase in others. On the other hand, a vaccination program via ``herd immunity'' might have a strictly beneficial effect on the risk of infection. 

Let $A$ denote the attributable effect of the treatment, defined to be the total difference between $Y$ and $\theta$:
\begin{equation}\label{eq: attributable}
A = \sum_{i=1}^N (Y_i - \theta_i).
\end{equation}
Our definition for $A$ generalizes that of \cite{rosenbaum2001effects} to allow for interference; if no interference is present, the two definitions are equivalent. Our inferential goal is a one-sided confidence interval lower bounding $A$. If this lower bound on $A$ is large, it implies that the observed treatment had a large effect on the outcomes. 

Let $G$ denote a network of observed pre-treatment social interactions between the units. This snapshot of observed interactions might be only a crude proxy for the actual social dynamics. Hence, we will not use $G$ to make explicit assumptions on the influence between units. Instead, $G$ will be used to choose a test statistic. Our motivation is robustness to model error.  If $G$ turns out to be a poor proxy, the method will lose power but not correctness, so that any significant findings will still be valid.

\section{Constructing a Confidence Interval for $A$} \label{sec: methods}

In this section, we present three methods for estimating one-sided confidence intervals that upper bound $\sum_i \theta_i$, which by \eqref{eq: attributable} is equivalent to a lower bound on the attributable effect $A$. In Section \ref{sec: CLT}, a t-test based asymptotic confidence interval is presented for count-valued outcomes, i.e., when $\theta$ and $Y$ are nonnegative integers.  In Section \ref{sec: wb}, a non-asymptotic estimate is presented for the special case of binary outcomes, which is then extended in Section \ref{sec: ws} to utilize the observed network $G$.

\subsection{T-test Based Asymptotic Confidence Interval} \label{sec: CLT}

Suppose that the entries of $\theta$ are actually observed for the $N-L$ untreated units. Assuming that these units are sampled without replacement, it is well known \cite{thompson2012sampling} that an unbiased point estimate for $\bar{\theta} = N^{-1} \sum_i \theta_i$ is given by the sample average $\hat{\theta}$,
\[\hat{\theta} = \frac{1}{N-L} \sum_{i: X_i = 0} \theta_i.\]
Under certain conditions, $\hat{\theta}$ is asymptotically normal, in which case an asymptotic $(1-\alpha)$ confidence upper bound for $\bar{\theta}$ is given by
\begin{equation} \label{eq: ideal CI}
\hat{\theta} + t_\alpha \sqrt{\left(\frac{L}{N}\right) \frac{\hat{\sigma}^2}{N-L}},
\end{equation}
where $\hat{\sigma}^2$ is the estimated variance, 
\begin{align*}
\hat{\sigma}^2 &= \frac{1}{N-L-1} \sum_{i:X_i = 0} (\theta_i - \hat{\theta})^2, 
\end{align*} 
and where $t_\alpha$ is the $\alpha$-critical value of a $t$ distribution with $N-L-1$ degrees of freedom. 

In our setting, $\theta$ is not actually observed, and hence \eqref{eq: ideal CI} cannot be evaluated. Let us assume that Assumption \ref{as: monotone} holds, and also that $\theta$ is restricted to the set of nonnegative integers, so that $0 \leq \theta \leq Y$ and $\theta \in \mathbb{Z}^N$. Then an upper bound to the unknown value of \eqref{eq: ideal CI} can be found by solving the following optimization problem:
\begin{align} \label{eq: conservative CI}
\max_{\theta \in \mathbb{Z}^N} &\quad  \hat{\theta} + t_\alpha\sqrt{\left(\frac{L}{N}\right) \frac{\hat{\sigma}^2}{N-L}} \\
\nonumber \st & \quad  0 \leq \theta_i \leq Y_i \textrm{ for all } i,
\end{align}
which equals the highest value of \eqref{eq: ideal CI} over all possible values of $\theta$. A polynomial-time solution method for this optimization problem is described in Appendix \ref{appendix: CLT}. 

\begin{example}
It may seem counterintuitive that \eqref{eq: conservative CI} may be maximized by $\theta$ smaller than $Y$. To illustrate that this may be possible, let $L=20$, $N=25$, and let the entries of $Y$ equal $(10, 10, 10, 11, 11)$ for the untreated units. Using \eqref{eq: ideal CI} while letting $\theta = Y$ gives a 95\% upper bound of $10.9$. On the other hand, letting $\theta$ equal $(0, 10, 10, 11, 11)$ for the untreated units gives an upper bound of $12.4$, achieving the optimal value of \eqref{eq: conservative CI}.
\end{example}

As with any t-test, by using \eqref{eq: conservative CI} we are implicitly assuming that $\hat{\theta}$ satisfies a central limit theorem.  Equivalently, we may instead state that one of two alternatives must be true: either \eqref{eq: conservative CI} gives a correct confidence interval, or the $\alpha$-quantile of $\hat{\theta}$ (after studentization) is greater than $t_\alpha$, which for large $N-L$ and $L$ roughly equates to $\theta$ having heavy tails.\footnote{for example, \cite[Th. 1.1]{bloznelis1999berry} implies that $\left(N^{-1} \sum_i |\theta_i^3|\right) \cdot\big(N^{-1} \sum_i \left( \theta_i - \bar{\theta} \right)^2\big)^{-3/2}$ must be large. }

We remark that bootstrapping the untreated entries in $Y$ will not compute a confidence interval for $\hat{\theta}$, since in general $\theta \neq Y$. However, the bootstrap may be still useful as a distributional check, testing whether \eqref{eq: ideal CI} is valid for the point hypothesis $\theta = Y$.



\subsection{Non-asymptotic Confidence Interval for Binary Outcomes} \label{sec: wb}

For binary-valued outcomes, a non-asymptotic one-sided confidence interval for $\sum_i \theta_i$ can be computed. This can be done by a process known as ``inverting a test statistic''\footnote{In practice, inverting a test statistic to produce a confidence interval can potentially result in unstable behavior when the underlying assumptions are violated \cite{gelman2011blog}. While we do not recommend our methods when Assumption \ref{as: monotone} is violated, they do not suffer from this behavior. This is because \eqref{eq: exact} will always have at least one feasible solution, $\theta=0$.}. Let $W(X;\theta)$ denote a test statistic of $X$ that is parameterized by the unknown $\theta$. Let $w_\alpha(\theta)$ denote the $\alpha$-quantile of $W(X;\theta)$, defined by
\begin{equation}\label{eq: critical value}
\mathbb{P}\left( W(X;\theta) \leq {w}_\alpha(\theta) \right) = \alpha.
\end{equation}
While $\theta$ is unknown, we know two constraints on its value. First, we know that $\theta \leq Y$, by Assumption \ref{as: monotone}. Second, we know that $W(X;\theta) \leq {w}_\alpha(\theta)$ with probability $\alpha$, by \eqref{eq: critical value}. Hence, to upper bound $\sum_i \theta_i$ with probability $\alpha$, we can find the $\theta$ which maximizes $\sum_i \theta_i$ while satisfying these constraints. That is, we can solve the optimization problem
\begin{align} \label{eq: exact}
\max_{\theta \in \{0,1\}^N} & \sum_{i=1}^N \theta_i \\
\nonumber \st &\hskip.2cm W(X;\theta) \leq {w}_\alpha(\theta) \\
\nonumber & \hskip.2cm \theta_i \leq Y_i \textrm{ for all } i.
\end{align}
It can be seen that $\eqref{eq: exact}$ includes all non-rejected hypotheses, thus finding a one-sided confidence interval for $\sum_i \theta_i$. 

We will use the test statistic $\Wb$, defined as
\[ \Wb(X;\theta) = \sum_{i=1}^N X_i \theta_i.\]
It can be seen that $\Wb(X;\theta)$ is generated by sampling $L$ entries from $\theta$ without replacement, so that $\Wb(X;\theta)$ is a $\operatorname{Hypergeometric}(\sum_i \theta_i, N - \sum_i \theta_i, L)$ random variable. As a result, the optimization problem \eqref{eq: exact} is easily computable for $W = \Wb$, and we describe a solution method in Appendix \ref{appendix: wb}. This method was originally presented in \cite[Appendix]{rosenbaum2001effects}, but for the case of no interference. 

\paragraph{Weaker Assumption} We present a weaker assumption than Assumption \ref{as: monotone}, which may be applicable when the treatment effect is not strictly nonnegative:
\begin{assumption}[Aggregate Monotonicity for the Untreated] \label{as: group monotone}
\[\sum_{i: X_i=0} \theta_i \leq \sum_{i:X_i=0} Y_i.\]
\end{assumption}
Unlike Assumption \ref{as: monotone}, which requires the treatment effect to be nonnegative for every individual, Assumption \ref{as: group monotone} only restricts the sum of the treatment effect over those units which did not receive treatment.

To upper bound $\sum_i \theta_i$ under Assumption \ref{as: group monotone}, we can solve a modification of \eqref{eq: exact},
\begin{align} \label{eq: weaker bound}
\max_{\theta \in \{0,1\}^N} & \sum_{i=1}^N \theta_i \\
\nonumber \st &\hskip.2cm W(X;\theta) \leq {w}_\alpha(\theta) \\
\nonumber & \hskip.2cm \sum_{i:X_i=0} \theta_i \leq \sum_{i:X_i=0} Y_i,
\end{align}
where we have replaced the constraint $\theta \leq Y$ by Assumption \ref{as: group monotone}. Details of the solution method for $W = \Wb$ are given in Appendix \ref{appendix: wb}.


\subsection{Using the observed network $G$} \label{sec: ws}

We extend the approach of Section \ref{sec: wb} to handle a new statistic $\Ws$, which utilizes the observed network $G$. This statistic will have power to detect treatment effects that spill over from treated units to their untreated neighbors. 

Let $\Ws$ be given by
\begin{align}
\nonumber \Ws(X;\theta) &= \frac{1}{L} \Wb(\tilde{X}; \theta) \\
\nonumber &= \frac{1}{L} \sum_{i=1}^N \tilde{X}_i \theta_i, 
\end{align}
where $\tilde{X}$ is a smoothed version of $X$, so that each entry in $\tilde{X}$ is a weighted average of nearby entries in $X$. More precisely, let $\tilde{X}$ equal
\[ \tilde{X} = X^T K,\]
where the smoothing matrix $K \in \mathbb{R}^{N \times N}$ is given by
\begin{equation} \label{eq: K}
 K_{ij} = \begin{cases} \frac{1}{Z_j} \exp(-d_{ij}^2/\sigma_K^2) & \mbox{if } d_{ij} \leq d_{\max,K} \\
0 & \mbox{otherwise,} \end{cases}
\end{equation}
where $d_{ij}$ denotes the distance between units $i$ and $j$ in $G$; where $d_{\max, K} \geq 0, \sigma_K > 0$ are shape parameters; and where $Z_j$ denotes a normalizing constant
\[Z_j = \sum_{i: d_{ij}\leq d_{\max, K}} \exp(-d_{ij}^2/\sigma_K^2),\]
chosen so that the columns sum to one, making each element of $\tilde{X}$ a weighted average of elements in $X$. 

Because each entry of $\tilde{X}$ is a weighted average, units that are close to treated units will have high values in $\tilde{X}$, even if they are not treated themselves. This will give $\Ws$ power to detect spillovers. However, unlike $\Wb$, exact solution of \eqref{eq: exact} is not computationally feasible for $W = \Ws$. In Appendix \ref{appendix: ws}, \eqref{eq: upper lagrange CLT} gives a relaxation of \eqref{eq: exact} that can be efficiently solved when the outcomes are binary-valued, yielding a asymptotically conservative  estimate of $A$ under Assumption \ref{as: monotone}.

\section{Data and Simulation Examples} \label{sec: simulation}
 
In this section, we present data and simulation examples to exhibit the performance of the methods described in the previous section. In Section \ref{sec: worms}, the estimator \eqref{eq: conservative CI} is used to analyze a primary school deworming experiment presented in \cite{miguel2004worms}. In Section \ref{sec: facebook experiment}, the Facebook election experiment of \cite{bond201261} is analyzed using the test statistic $\Wb$. In Section \ref{sec: spatial simulation}, simulated experiments are used to evaluate the performance of the test statistic $\Ws$. 


\subsection{Analysis of \cite{miguel2004worms}} \label{sec: worms} 

\cite{miguel2004worms} describes a primary school deworming project that was carried out in 1998 in Busia, Kenya, in order to reduce the number of infections by parasitic worms in young children. We restrict analysis to $N=50$ schools in a high infection area of Busia, which were divided into 2 equal-sized groups. Schools in group 1 received free deworming treatments beginning in 1998, while group 2 did not. Students were surveyed in 1999, and substantially fewer infections were found in the treatment-eligible pupils in group 1 compared to group 2, with 141 and 506 infections respectively. It is believed that the number of infections in each schools was affected not only by its own treatment status, but also that of other schools as well. This is because students that received the deworming treatment were susceptible to re-infection by infected students. 

To demonstrate the estimator given by \eqref{eq: conservative CI} on this experiment, we will assume that treatment was assigned by sampling without replacement\footnote{Groups 1, 2, and 3 (with group 3 excluded from the 1999 survey) were actually assigned by dividing the schools into administrative subunits, listing them in alphabetical order, and assigning every third school to the same group.}, and that all missing values in the data are ignorable. We also assume that the deworming treatment never increases the risk of infection, either to its direct recipient or to others. Under these assumptions, we solve a variant of \eqref{eq: conservative CI} as discussed in Appendix \ref{appendix: CLT}. The resulting estimates are that with 95\% confidence, the number of infections that would have occurred if all schools received deworming is upper bounded by $347$, and the number of infections that would have occurred if no schools received deworming is lower bounded by $829$. These estimates may well be conservative, as no spatial information was used. However, they are not vaccuous; the one-sided confidence intervals are equal to those given by a regular t-test, which requires a much stronger assumption of no interference between schools, and an identical assumption regarding the asymptotic normality of $\hat{\theta}$. 

\subsection{Election Day Facebook Experiment} \label{sec: facebook experiment}

Using the reported counts for each treatment/outcome combination for the Facebook experiment of \cite{bond201261}, we may estimate the attributable effect $A$ by solving \eqref{eq: exact} or \eqref{eq: weaker bound} for $W = \Wb$. In both cases, the resulting 95\% confidence interval for $A$ equals $[1199323, \,  \infty)$, implying that the usage of profile pictures caused at least 1,199,323 users to click ``I voted'', when they would not have done so otherwise. This equals 2.0\% of the treated population, matching the estimate of \cite{bond201261} which assumed no interference.


As the solutions to \eqref{eq: exact} and \eqref{eq: weaker bound} are the same, our estimate of $A$ is valid under either Assumption \ref{as: monotone} or Assumption \ref{as: group monotone}. Possibly, some individuals may have been discouraged from voting by seeing the profile picture of a Facebook friend (for example, perhaps due to a negative relationship), which would violate Assumption \ref{as: monotone}. Assumption \ref{as: group monotone} allows for this possibility, since no restrictions are made on the effects of treatment on the treated.

\subsection{Simulated Study} \label{sec: spatial simulation}

In settings where spillover effects are large, the statistic $\Ws$ may outperform $\Wb$ by identifying clusters of outcomes that were caused by the treatment. To demonstrate this behavior, we ran simulations in which treatments resulted in higher probabilities of positive outcomes not only for the treated units, but also for those nearby as well. We explored a range of scenarios, varying the number of treatments and their spatial separation, the spillover radius of the treatment effect, the counterfactual $\sum_i \theta_i$, and also the choice of kernel matrix $K$. We found that estimates using $\Ws$ were most accurate and robust to choice of $K$ when the treatments resulted in many well-separated clusters of positive outcomes; in particular, increasing the number of treatments or their potency could could actually decrease accuracy, by causing treatment effects to ``run into each other''.

\paragraph{Description of Simulated Experiments}

In each simulation, $N$ units were placed on a uniformly spaced $\sqrt{N} \times \sqrt{N}$ grid. Sampling with replacement was used to select units $j_1,\ldots,j_L$ for treatment, and auxiliary binary variables $Z_1,\ldots,Z_L$ were generated with distribution $\operatorname{Bernoulli}(1/2)$. For $i=1,\ldots,N$, each counterfactual outcome $\theta_i$ was a $\operatorname{Bernoulli}(p_0)$ random variable, and each observed outcome $Y_i$ equaled $1$ if $\theta_i=1$, and otherwise equaled a $\operatorname{Bernoulli}(P_i)$ random variable, where the probability $P_i$ of having outcome $Y_i=1$ due to treatment was given by
\begin{align} \label{eq: P_i}
P_i & = 1 - \prod_{\ell=1}^L (1 - h(i, j_\ell))^{Z_\ell},
\end{align}
where $h$ denotes a truncated gaussian, 
\begin{align} \label{eq: h}
h(i,j) = \begin{cases} 0 & \textrm{if } d_{ij} > d_{\textrm{max}, h} \\
\min\left(1 , C \exp\{-d_{ij}^2/ \sigma_h^2\}\right) & \textrm{otherwise}, \end{cases}
\end{align}
where $d_{ij}$ denotes distance between units $i$ and $j$ on the grid, and where $d_{\textrm{max},h}, C,$ and $\sigma_h$ are shape parameters. In words, \eqref{eq: P_i}-\eqref{eq: h} imply that each treatment $\ell$ has no effect if $Z_\ell=0$, and otherwise has an area effect that is independent of other treatments, i.e., each treatment $\ell$ for which $Z_\ell=1$  has probability $h(i,j_\ell)$  of independently causing unit $i$ to have outcome $Y_i = 1$.

For each experiment, estimation using $\Ws$ was computed by solving \eqref{eq: upper lagrange CLT}, which is a relaxation of \eqref{eq: exact} as discussed in Appendix \ref{appendix: ws}. In all simulations where the spillover effects were large, we note that $\Wb$ and \eqref{eq: conservative CI} gave nearly vacuous estimates, since they cannot detect spillovers.

\paragraph{Simulation Results}

\begin{figure}
\begin{center}
    \begin{subfigure}[t]{.48\textwidth} 
  	\includegraphics[width=3in]{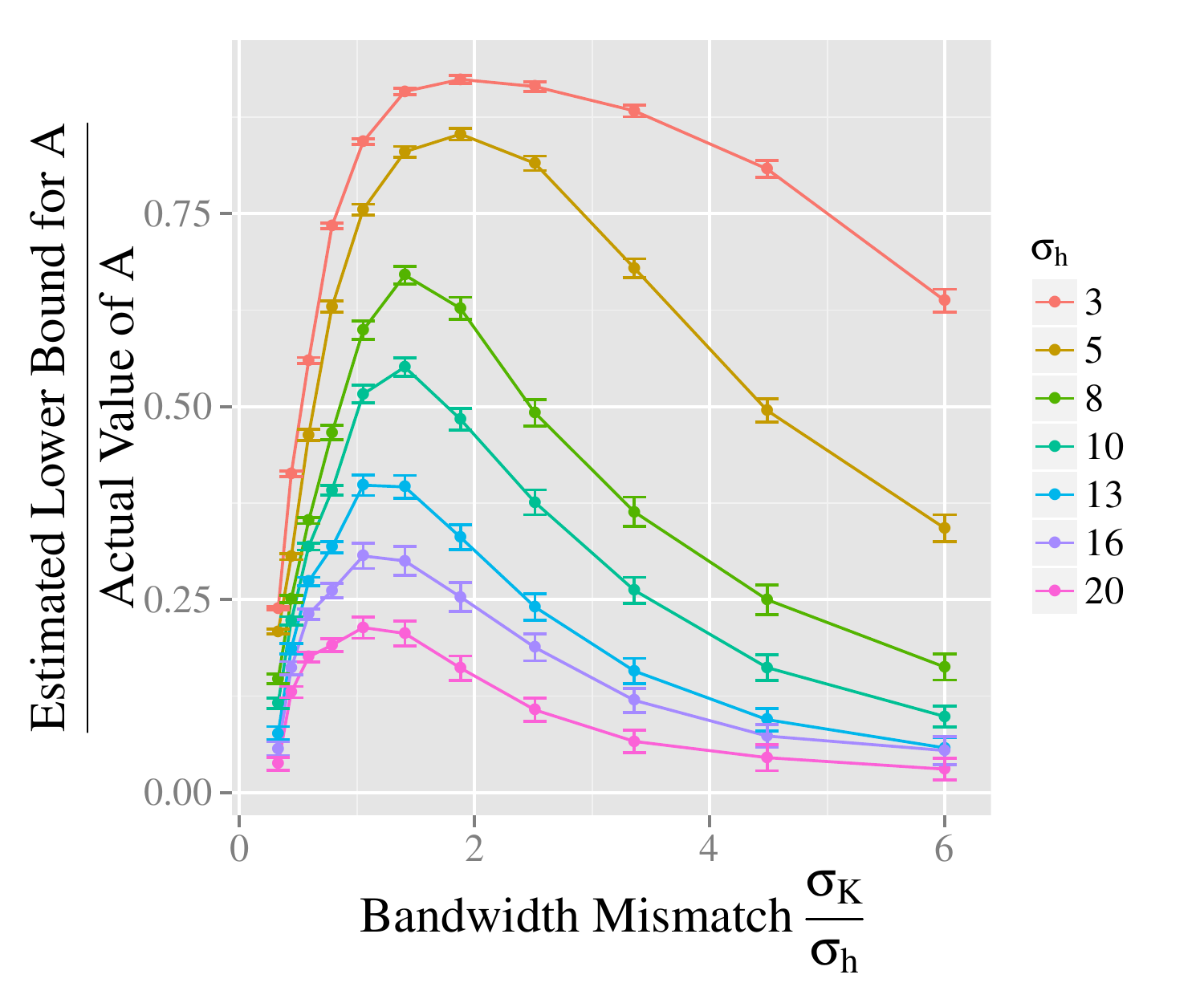}
        \caption{Estimation accuracy}
	\label{fig: sim1a}
    \end{subfigure} 	
    \begin{subfigure}[t]{.48\textwidth}
	\includegraphics[width=3in]{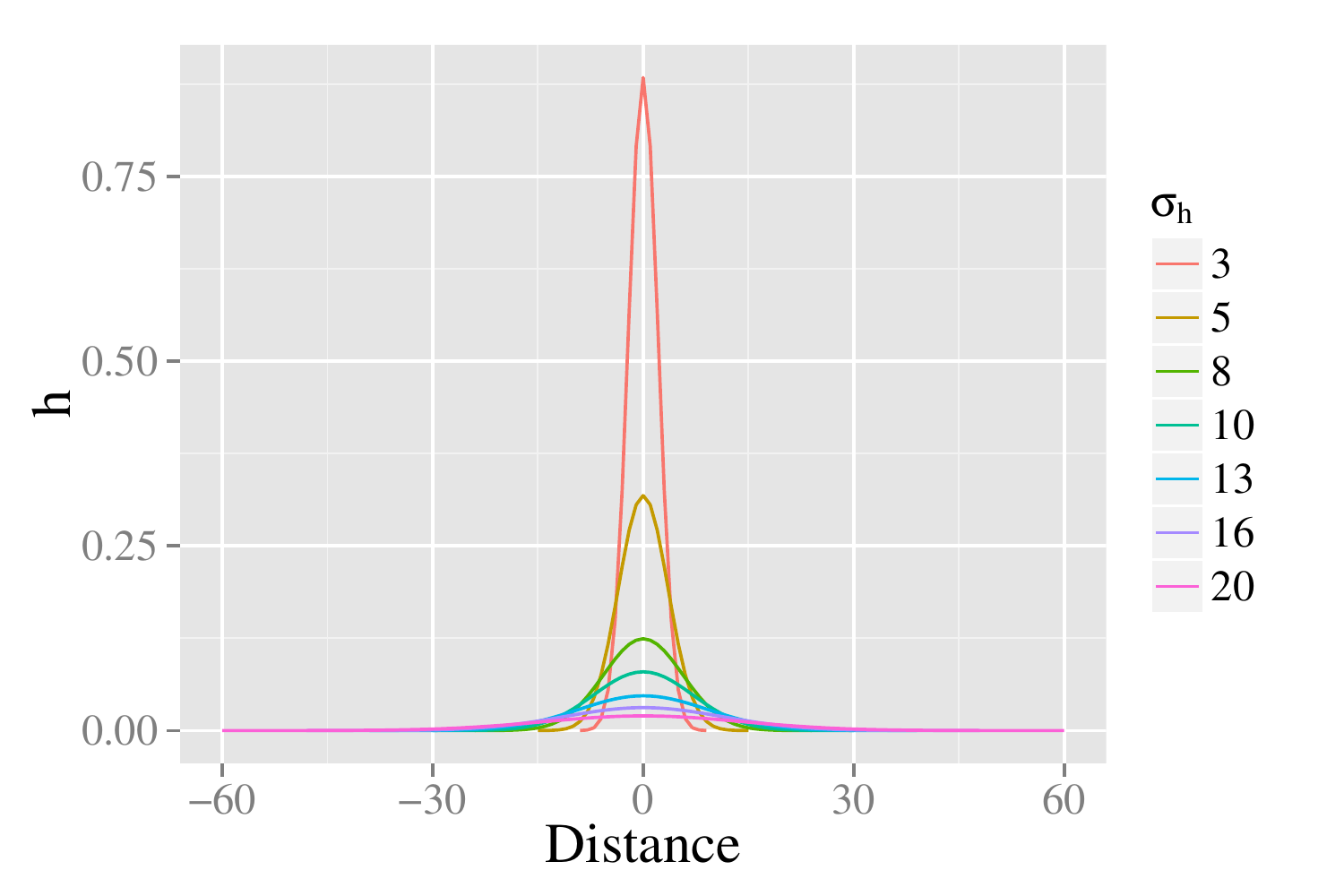}
        \caption{Various $h$ used in simulations}
	\label{fig: sim1b}
    \end{subfigure}
    \caption{Average accuracy (and standard errors) of estimated lower bound for $A$, for various choices of spillover function $h$ and mismatched smoothing matrix $K$. The spillover functions $h$, shown in (b), were chosen by varying the bandwidth $\sigma_h$ while keeping $A$ constant in expectation. $K$ was chosen to have a mismatched bandwidth $\sigma_K$ that was a multiple of the generative $\sigma_h$. 100 simulations per data point; examples of the simulations are shown in Fig. \ref{fig: sim1c}. \label{fig: sim1}}
\end{center}
\end{figure}

\begin{figure} 
 \centering
 \begin{subfigure}[t]{.3\textwidth}
	\includegraphics[width=2.063in]{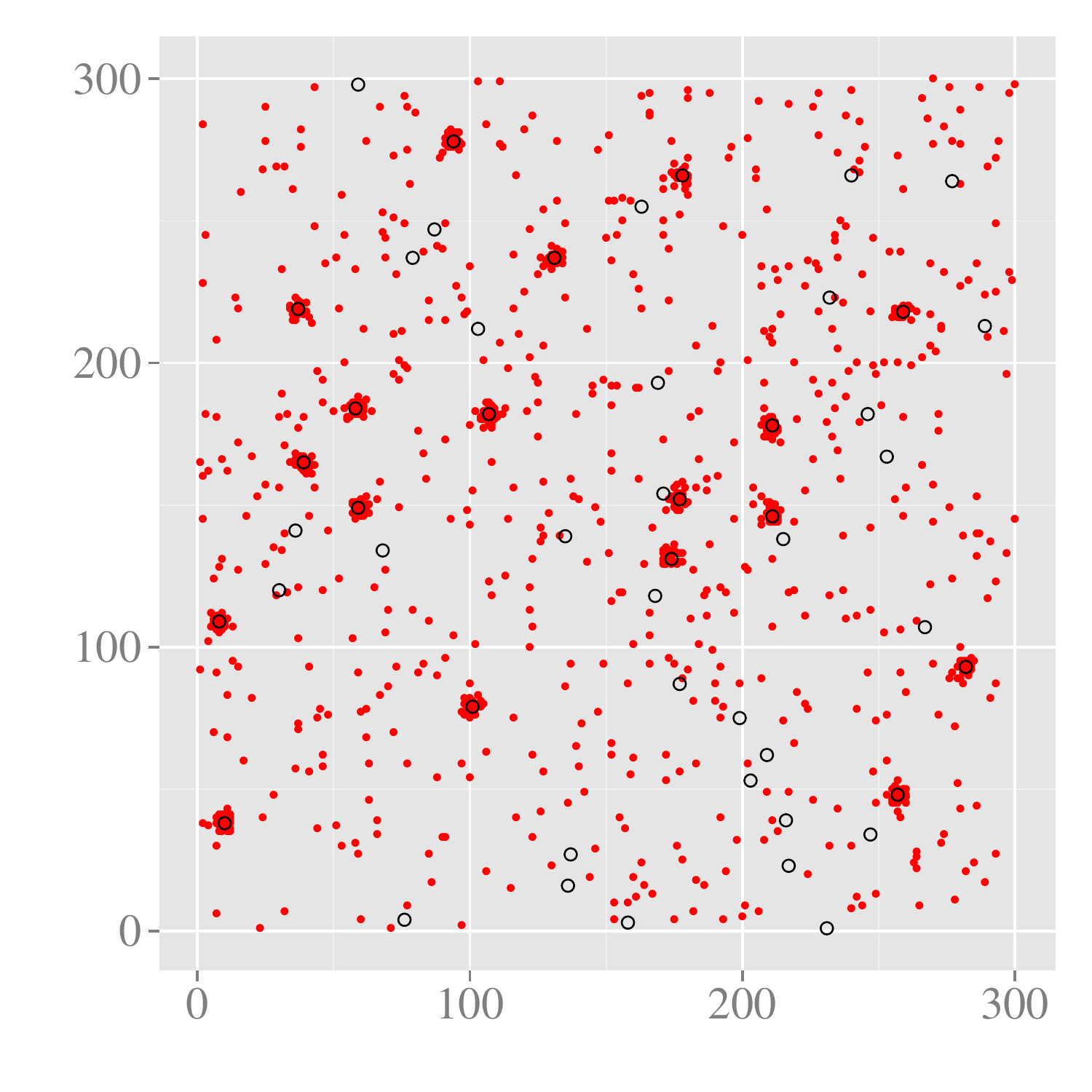}
	\caption{$\sigma_h = 3$; highly localized}
 \end{subfigure}
 \begin{subfigure}[t]{.3\textwidth}
	 \includegraphics[width=2.063in]{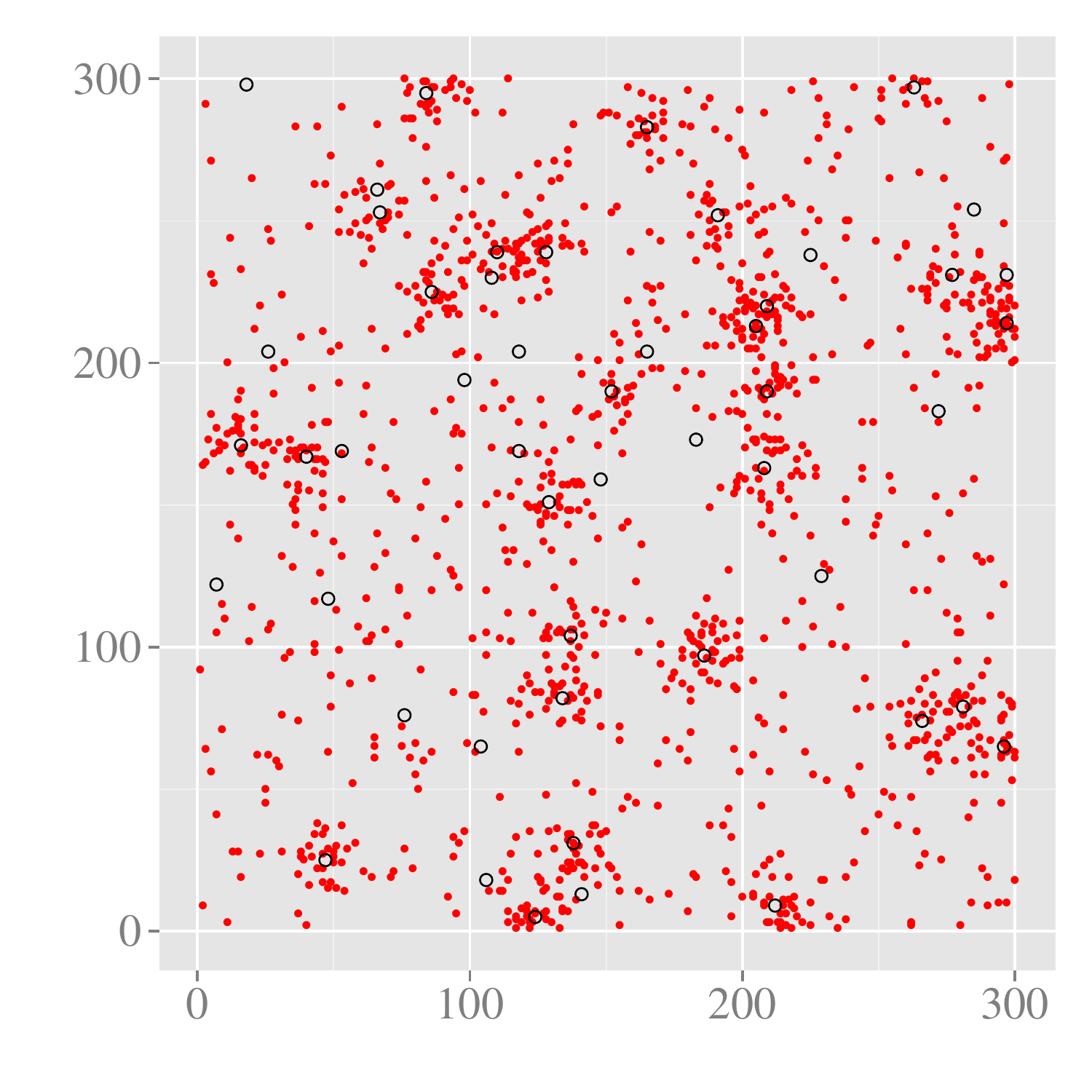}
	\caption{$\sigma_h = 10$; medium localization}
 \end{subfigure}
 \begin{subfigure}[t]{.3\textwidth}
	 \includegraphics[width=2.063in]{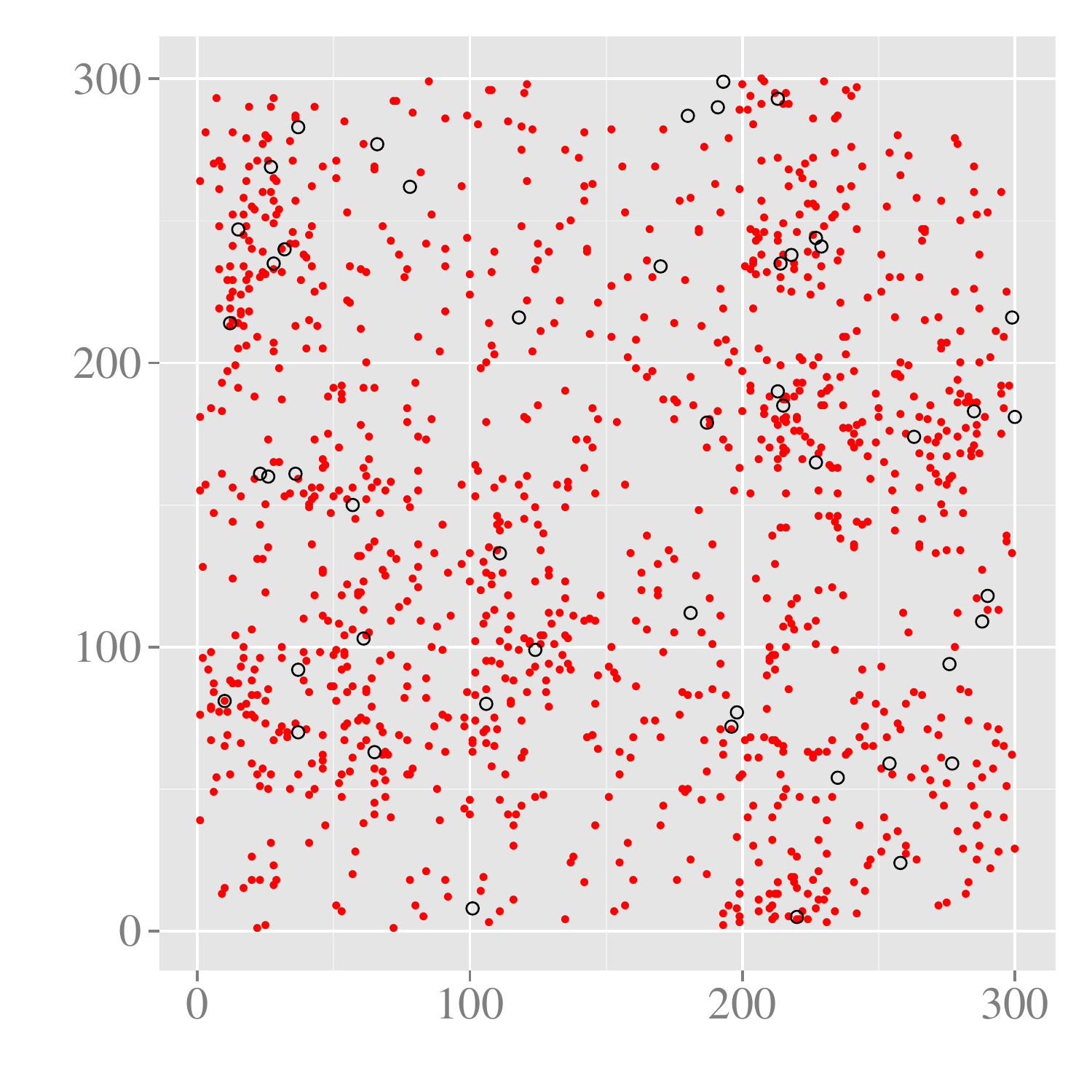}
	\caption{$\sigma_h = 20$; diffuse effects}
 \end{subfigure}
 \caption{Examples of simulated experiments used to generate Fig. \ref{fig: sim1}, in which the spillover function $h$ was varied while the expectation of $A$ was held constant. $N=90,000$ units were placed on a $300 \times 300$ grid. Black circles denote treated units ($L=50$), red dots denote units with outcome 1. Treatment effects were large; on average, each treatment caused 12.5 outcomes, and $\sum_i Y_i = 1225$ and $\sum_i \theta_i = 600$ in expectation. \label{fig: sim1c}}
\end{figure}

Figure \ref{fig: sim1a} shows estimation performance as a function of the generative $h$ and the assumed kernel $K$. To construct this figure, 7 different choices for $h$ were used, in which $\sigma_h$ and $C$ were adjusted so that the degree of localization of the treatment effect was varied while $A$ was kept constant in expectation. These choices for $h$ are shown in Figure \ref{fig: sim1b}, with examples of the simulated outcomes shown in Figure \ref{fig: sim1c}. The assumed kernel $K$ was varied by ranging the bandwidth parameter $\sigma_K$ used in \eqref{eq: K} from $\sigma_h/3$ to $6 \sigma_h$. In all cases, performance eventually decreased for large $\sigma_K$, suggesting that the choice of $K$ should reflect knowledge about the anticipated treatment effect. For localized effects (i.e., small $\sigma_h$), the estimates were more accurate, and allowed for the bandwidth of $K$ to be chosen many times larger than $\sigma_h$. For diffuse effects (i.e., large $\sigma_h$), estimates were highly conservative and more sensitive to the choice of $K$. These results suggest that estimation using $\Ws$ may require spatial separation between treated units, so that the effects can be localized to their source. 

\begin{figure}
\begin{center}
    \begin{subfigure}[t]{.48\textwidth} 
  	\includegraphics[width=3in]{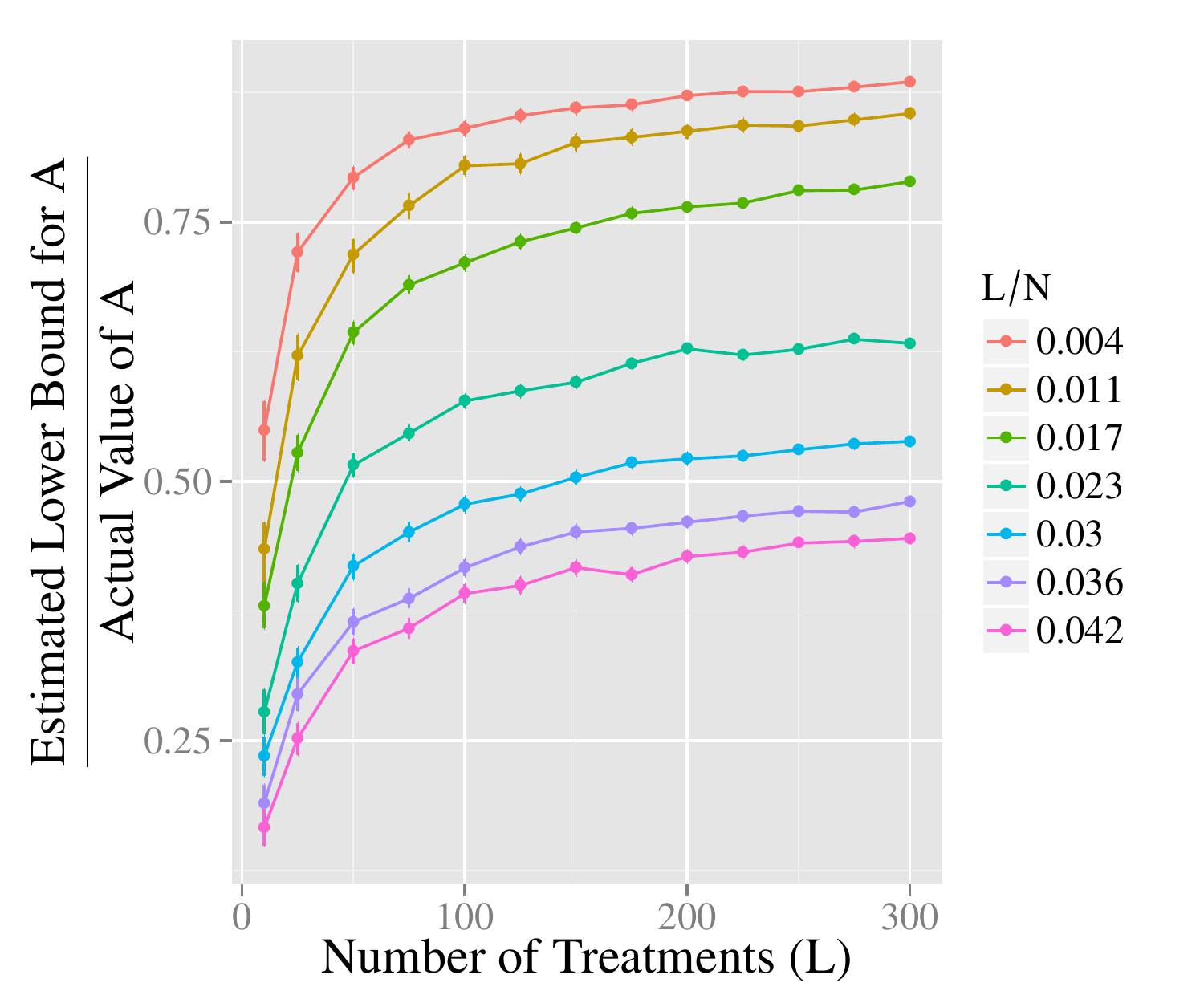}
    \end{subfigure} 	
\caption{Average estimation accuracy (and standard errors) using $\Ws$ with smoothing matrix $K$ matched to the generative $h$, while varying the number of treatments $L$ and their spatial density $L / N$. 400 simulations per data point; examples of the simulations are shown in Fig. \ref{fig: sim2c} \label{fig: sim2a}}
\end{center}
\end{figure}

\begin{figure} 
 \centering
 \begin{subfigure}[t]{.3\textwidth}
	\includegraphics[width=2.063in]{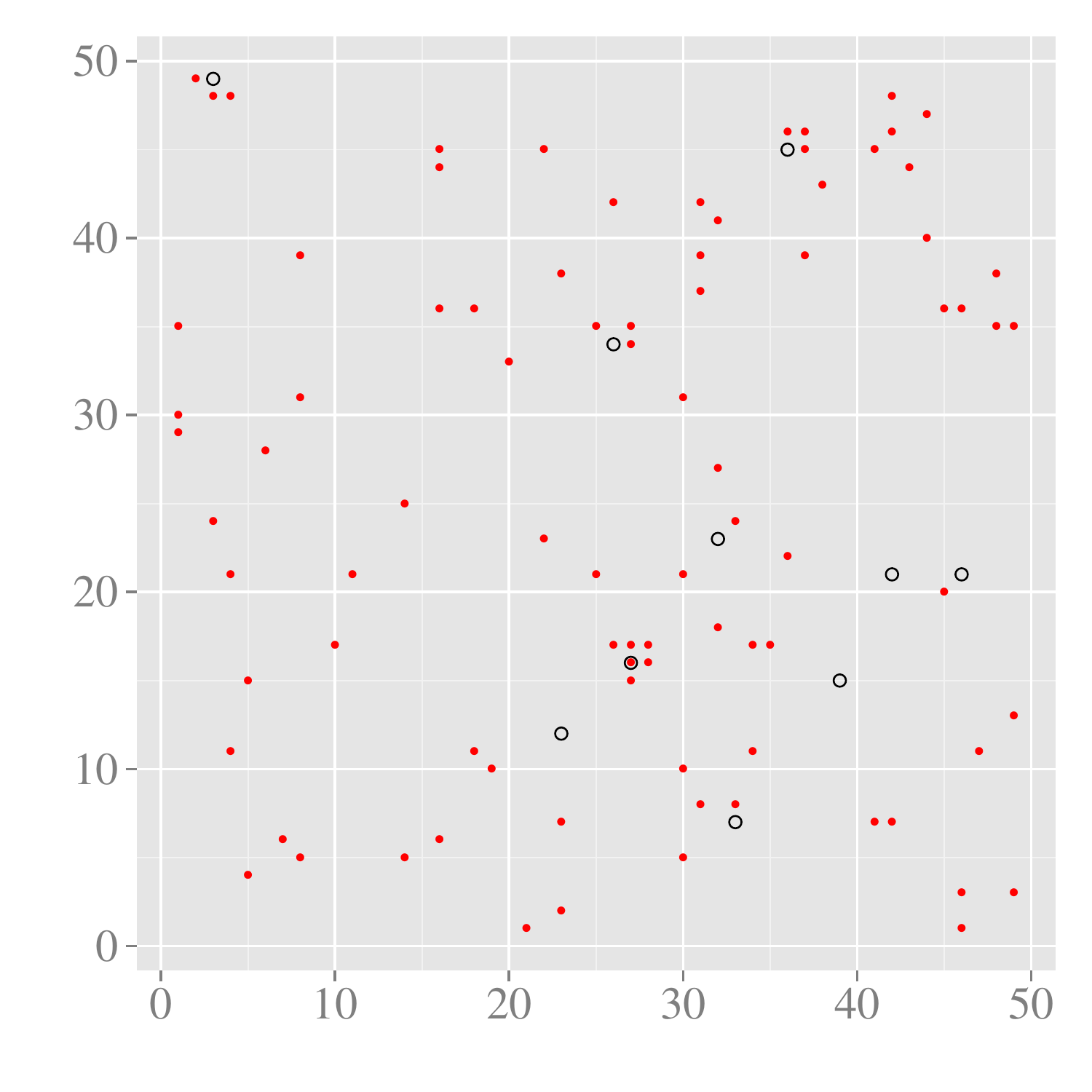}
	\caption{$L = 10, L/N = 0.04$\\  \phantom{abcde} $(50 \times 50$ grid)}
 \end{subfigure}
 \begin{subfigure}[t]{.3\textwidth}
	 \includegraphics[width=2.063in]{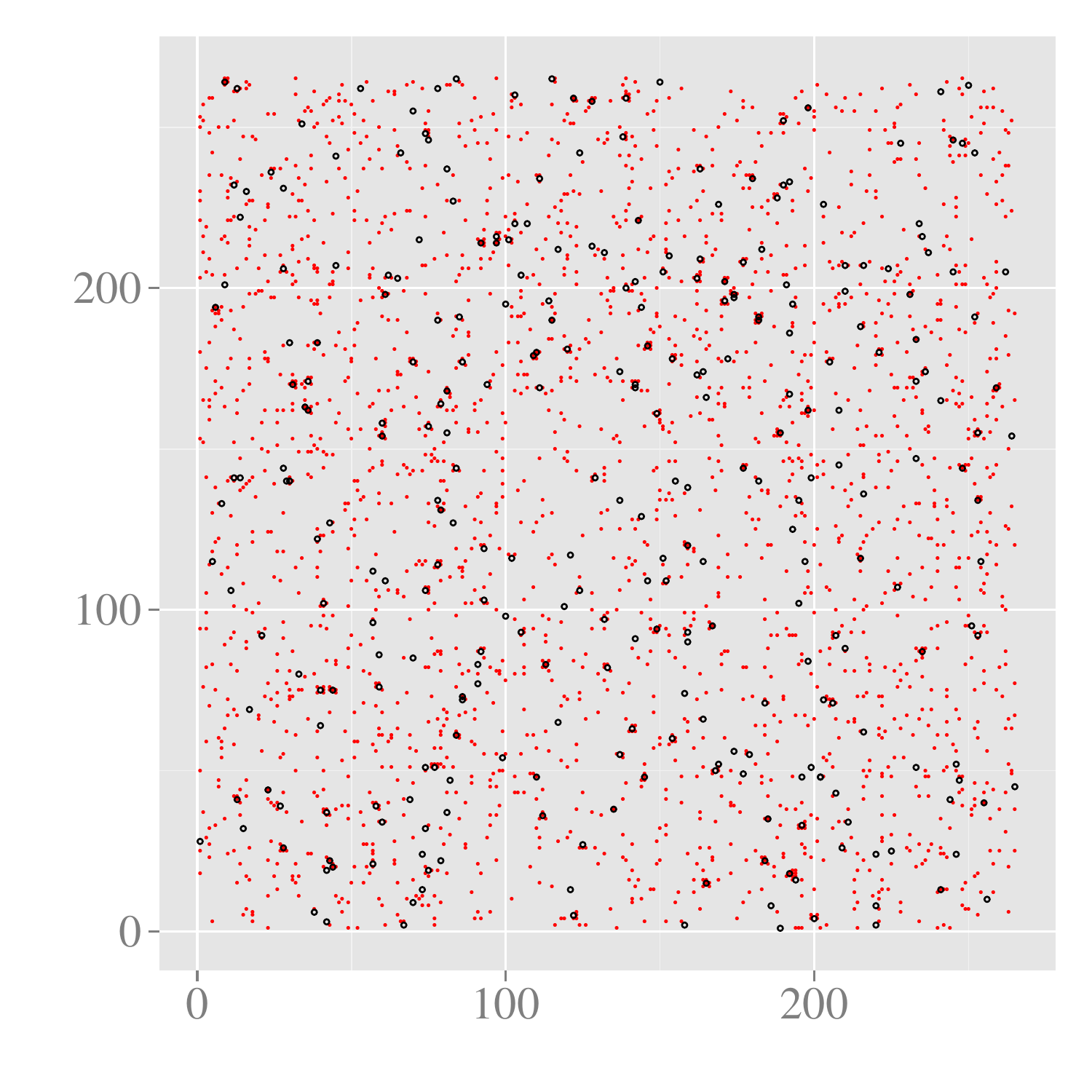}
	\caption{$L = 300, L/N = 0.04$ \phantom{abcde} ($265 \times 265$ grid)}
 \end{subfigure}
 \begin{subfigure}[t]{.3\textwidth}
	 \includegraphics[width=2.063in]{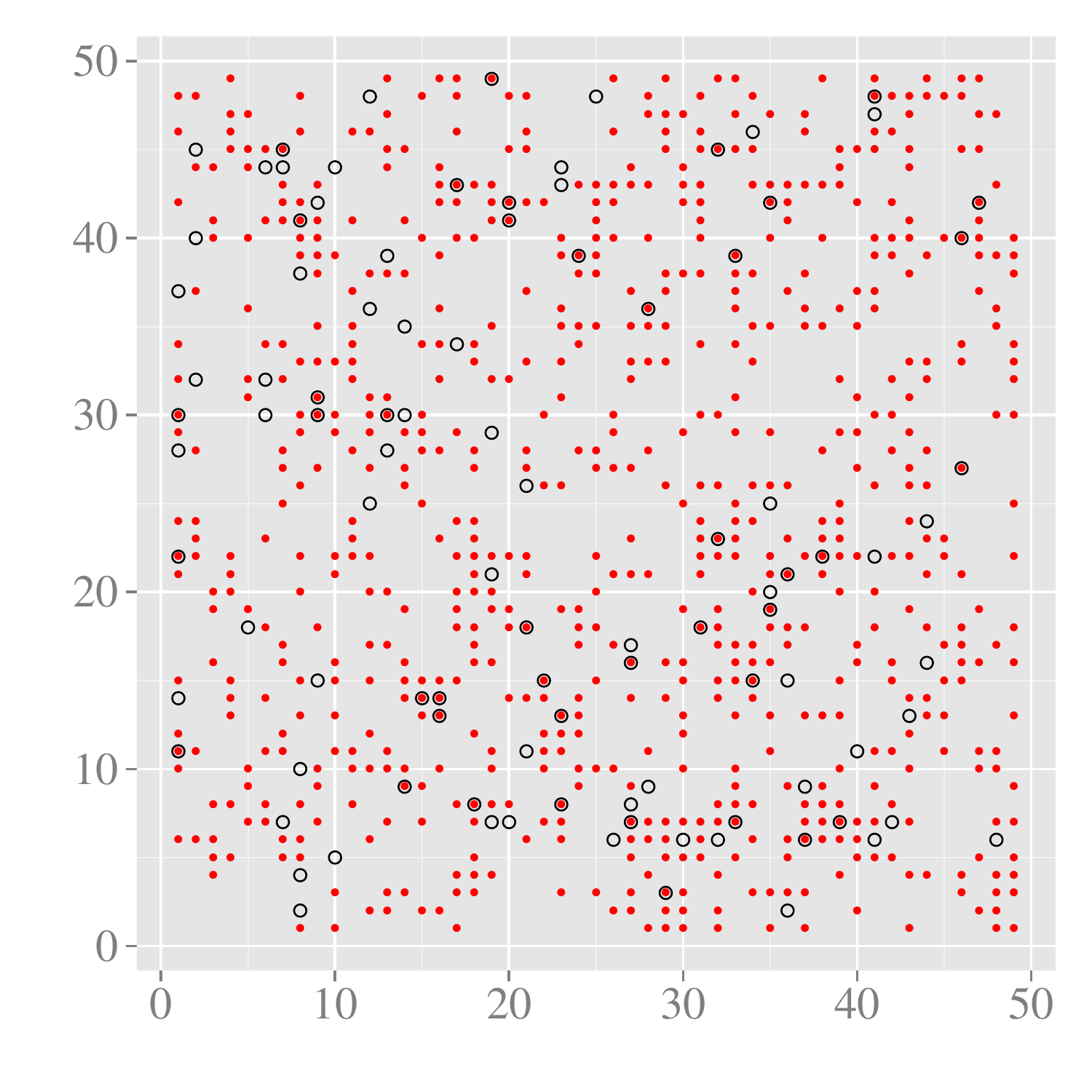}
	\caption{$L = 100, L/N = 0.36$ \phantom{abcde}($50 \times 50$ grid)}
 \end{subfigure}
 \caption{Examples of simulations used to generate Figure \ref{fig: sim2a}. (a) and (b) show low density treatments on small and large grids, while (c) shows high density treatments on a grid of equal size to (a). Estimation accuracy was best for (b), then (a), and worst for (c). Each treatment caused 1.5 outcomes on average, and $\sum_i \theta_i / \sum_i Y_i = 0.7$ in expectation. \label{fig: sim2c}}
\end{figure}

Figure \ref{fig: sim2a} shows average estimation performance as a function of the number of treatments $L$, and also their spatial density $L/N$, which was controlled by varying the grid size $N$. We found that increasing with the number of treatments improved accuracy, while increasing the spatial density of treatments worsened it. As a result, increasing $L$ while keeping $N$ fixed could decrease accuracy,  due to the diminished spatial separation between the treatments.  Examples of the simulations used are shown in Figure \ref{fig: sim2c}. 

\begin{figure}
\begin{center}
    \begin{subfigure}[t]{.48\textwidth} 
  	\includegraphics[width=3in]{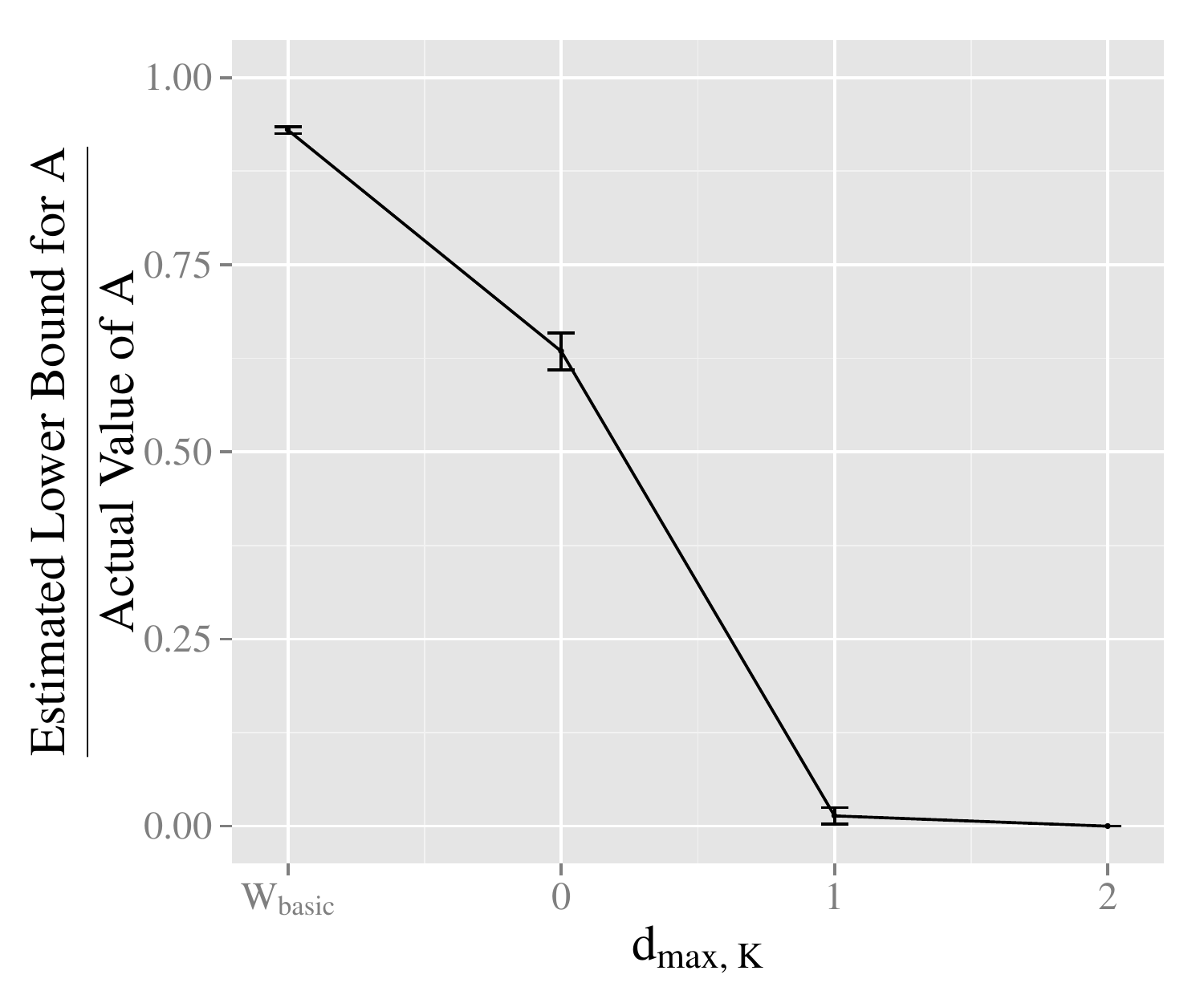}
    \end{subfigure} 	
\caption{Estimation accuracy (average performance and standard errors) in spatial experiments in which the treatment had only a direct effect (i.e., no spillovers). Estimation either used $\Wb$, or used $\Ws$ with $d_{\max,K}$ varied between $0$ (no spillovers assumed) to $2$ (spillovers up to distance 2 assumed). Experiments involved $N=90,000$ units placed on a $300 \times 300$ grid, with $L=50$ treatments. $\sum_i \theta_i = 600$ and $\sum_i Y_i = 625$ in expectation. 100 simulations per data point. \label{fig: sim3a}}
\end{center}
\end{figure}

Figure \ref{fig: sim3a} shows the average estimation performance when $d_{\max, h} =0$, meaning that the simulated treatments had no spillovers. The estimated lower bound on $A$ was produced either by inverting $\Wb$, or by inverting $\Ws$ with $d_{\max, K} = 0,1,2$; the parameter $d_{\max,K}$ can be interpreted as an assumption on the maximum distance between a treated unit and its spillover. Estimation using $\Wb$ was most accurate; on average, the estimated lower bound on $A$ was $93\%$ of the true value. Estimation using $\Ws$ was less accurate, ranging from $63\%$ of the true value when $d_{\max,K}=0$ to the trivial lower bound of zero when $d_{\max,K}=2$. These results reinforce that $K$ should reflect knowledge of the anticipated treatment effect, and that $\Wb$ may perform better when spillovers are at zero or near-zero levels.

As expected, the coverage rates for the estimated 95\% one-sided confidence intervals were conservatively high. The highest frequency of violated confidence intervals was 3\%, which occurred when $L=10, L/N = 0.04$. Over all of the simulations, only 0.1\% of them resulted in a confidence interval which did not cover the true value of $A$.

\section{Discussion} \label{sec: discussion}

\paragraph{Applicability of $\Ws$}

The simulations of Section \ref{sec: spatial simulation} are stylized, and are mainly meant to show that in principle, it is possible to rigorously estimate spillovers without placing strong assumptions on the validity of the observed network $G$. However, the results also suggest that as a practical method, inverting the test statistic $\Ws$ may have limitations due to the following requirements:
\begin{enumerate}
\item The treatments should result in a large number of well-separated clusters of outcomes. If spillovers are non-existent or very small, $\Wb$ should be used instead.
\item The kernel smoothing matrix $K$ should be at least somewhat matched to the form of the spillovers. 
\end{enumerate}
How practical are these requirements? We would not expect the effects of single physical treatment, such as a coupon or advertisement, to resemble the simulations, in which as many as $12.5$ outcomes were caused per treatment. However, the condition $X_i=1$ need not represent a single physical treatment. Instead, it could mean administering the physical treatment to a subset of units in the vicinity of $i$. For example, the condition $X_i=1$ could signify that some percentage of all units within some distance to $i$ (or belonging to the same region as $i$) receive the physical treatment. In this manner, it may be possible to design experiments in which the outcomes tend to be clustered at some desired intensity. Additionally, the treatment vicinities corresponding to each unit may be used to guide the choice of the kernel smoothing matrix $K$. 

Cluster-randomized designs, such as the type described above, are likely to be more effective for investigating interference-based effects -- not only for $\Ws$, but for any other estimator as well. Assumption \ref{as: monotone} allows for a good deal of flexibility in the experiment design. For example, if a unit belonged to multiple vicinities that were selected for treatment, the experiment protocol could give the unit a higher probability of receiving the physical treatment, or limit the unit to the same probability as those units in a single treatment vicinity, or even disqualify the unit from treatment altogether, as all three design options are allowed under Assumption \ref{as: monotone}.

\paragraph{General Usage}

In this paper, we have considered the problem of estimating the attributable effect $A$ by a lower bound. Such a lower bound, if it is not vacuously conservative, may help in determining whether an experimental treatment had a practically significant effect. In returning only a lower bound, we are taking a conservative approach to the possibility of errors in the network or spatial model (or the lack of a model in \eqref{eq: conservative CI} and $\Wb$). We believe that a conservative approach to model misspecification will be desireable in some applications.

In addition to estimation of $A$, one might consider testing the hypothesis that $A = \sum_i (\theta_i - Y_i)$ equals zero. However, under Assumption \ref{as: monotone}, $A$ can equal zero only if $\theta = Y$, meaning that the treatment must have zero effect on each individual unit. As a definition of ``no effect'', this is far more restrictive than the hypothesis of zero average treatment effect, which allows for individual outcomes to change under treatment so long as the totals remain the same. For this reason, we recommend that significance tests should not assume Assumption \ref{as: monotone}. When interference is present, a better choice for significance testing might be to use the rank-based methods of \cite{rosenbaum2007interference}.

While we have focused on estimation of the attributable effect $A$, our methods can sometimes also be applied to estimate a version of the average treatment effect, which we define as follows. Let  $\theta^{\FT}$ denote the counterfactual outcomes under full treatment, i.e., the outcome if all units were treated and $X_i=1$ for all $i$. Let $\theta^{\FC} \equiv \theta$ denote the counterfactual under full control. One definition for the average treatment effect is
\[ ATE = \frac{1}{N} \sum_{i=1}^N (\theta_i^{\FT} - \theta_i^{\FC}),\]
which is the difference in outcomes between full treatment and full control, averaged over all units. As an example, in Section \ref{sec: worms} (and with further details in Appendix \ref{appendix: CLT}), we report an upper bound on $\sum_i \theta^{\FT}$ and a lower bound on $\sum_i \theta^{\FC}$ using \eqref{eq: conservative CI} for the data of \cite{miguel2004worms}, thus inducing a lower bound on the average treatment effect. For binary outcomes, it can be seen that solving \eqref{eq: exact} for $\Wb$ with $1-X$ in place of $X$ and $1-Y$ in place of $Y$ is equivalent to estimating a upper bound on $1 - \sum_i \theta_i^{\FT}$, which gives a lower bound on $\sum_i \theta^{\FT}$. In principle, \eqref{eq: upper lagrange CLT} for $\Ws$ can also be solved with $X$ and $Y$ transformed in the same manner. However, the runtime for inverting $\Ws$ for this problem will be prohibitively large if  $\sum_i (1 - Y_i) \gg \sum_i Y_i$, as was the case in the simulations. 
As a result, the performance of the relaxation \eqref{eq: upper lagrange CLT} under this transformation has not been investigated. 




\paragraph{Future directions and further analysis of \cite{miguel2004worms}}

In many settings, an observed network $G$ or spatial information might be only a crude proxy to the true underlying social mechanisms. We have shown that it is possible to rigorously use such information to improve estimates, without making unreasonable assumptions on the generative process. However, the proposed method needed high signal-to-noise for good performance, and it was not demonstrated on a real data set. For these reasons, usage of $\Ws$ should be regarded as proof-of-concept rather than recommended practice. 

As a possible direction for future work, we are investigating how the method of \eqref{eq: conservative CI} might be applied to the ``effective treatment'' estimator discussed in \cite[Sec. 2.4.3]{eckles2014design}. This estimator, also discussed in \cite{aronow2012estimating}, was shown in \cite[Thm 2.2]{eckles2014design} to reduce bias under Assumption \ref{as: monotone}, but currently requires a correctly specified exposure model to compute a confidence interval. As this is a very strong assumption, a conservative estimate similar to \eqref{eq: conservative CI} may be of interest.

We describe a special case of this estimator for which \eqref{eq: conservative CI} can be seen to apply, in the context of the deworming experiment of \cite{miguel2004worms}. We grouped $48$ of the $50$ schools into 16 triplets by order of distance, i.e., the closest three schools were grouped together, then the closest three out of the remaining schools, and so forth. The final 2 schools were removed from the analysis. We declared that a group of schools was treated if at least 2 schools in the group were treated (i.e., if they received the deworming treatment). The treated schools in the treated groups were declared to be selected. In this manner, 18 schools belonging to 8 treated groups were selected. Conditioned on the number of treated groups, and the number of selected schools in each group, the distribution of the 18 selected schools equals a two-stage sample \cite{thompson2012sampling}, in which the treated groups are selected by sampling without replacement, and then the selected schools are sampled within the treated groups. It follows by arguments similar to Section \ref{sec: CLT} that the average number of observed infections for the 18 selected schools is a conservatively biased point estimate for the per-school infections under full treatment. This value equaled $3.8$, implying an point estimate of $182$ for the total number of infections under full treatment. This is a $33\%$ reduction from the point estimate of $270$ that would result from an assumption of no interference, i.e., if all 24 treated schools were averaged.

To compute a confidence interval, in principle the method of \eqref{eq: conservative CI} can be applied to the selected schools, using the estimated variance of a two stage sample in place of $\hat{\sigma}$. While the small sample size of 8 groups likely invalidates the central limit theorem requirements of \eqref{eq: conservative CI}\footnote{We remark that the upper bound found this way for the deworming experiment was $297$. This is somewhat less than the estimate of $347$ found in Section \ref{sec: worms}, suggesting at least that the proposed approach will not be vacuously conservative.}, the approach may be applicable in a larger experiment, such as \cite{bond201261}. Also, we observe that the point estimate is reminiscent of a U-statistic, since it can be written as a function of all ${N \choose 3}$ school triplets and their respective treatments. This suggests further possibilities for new estimators.

In this preliminary analysis, the spatial information in \cite{miguel2004worms} was used to remove treated schools from consideration  if they were far from other treated schools. This improved the point estimate because such schools were more susceptible to reinfection. This is quite different from the simulations, where well-separated treatments gave the best estimates. We conjecture that both types of settings can arise in practice.

\section*{Appendices}

\appendix

\section{T-test Based Asymptotic Confidence Interval} \label{appendix: CLT}

\paragraph{Solution of \eqref{eq: conservative CI}} It can be seen that the objective function of \eqref{eq: conservative CI} is a function of $\hat{\theta}$ and $\hat{\sigma}^2$, and is increasing in the latter argument. Hence, the optimal $\theta$ will maximize $\hat{\sigma}^2$ over some level set of $\hat{\theta}$, which is equivalent to solving 
\begin{align} \label{eq: DP problem}
\max_{\theta \in \mathbb{Z}^N} &\quad \sum_{i:X_i=0} \theta_i^2 \\
\nonumber \st & \quad \sum_{i:X_i=0} \theta_i = c \\
\nonumber & \quad   0 \leq \theta_i \leq Y_i \textrm{ for all }i,
\end{align}
for some value of $c$. Since $c$ must be an integer between $0$ and $\sum_{i:X_i=0} Y_i$, we can solve \eqref{eq: DP problem} for all possible values of $c$, and then choose the solution that maximizes \eqref{eq: conservative CI}.



To solve \eqref{eq: DP problem}, let $n = N-L$ and let $i_1,\ldots,i_n$ sort the elements of $\{Y_i:X_i=0\}$ in descending order. It can be seen that \eqref{eq: DP problem} is maximized by letting $\theta_{i_1} = \min\left\{ c, Y_{i_1}\right\}$, and following the recursion
\begin{equation} \label{eq: DP easy solution}
\theta_{i_j} = \min\left\{ c - \sum_{k=1}^{j-1} \theta_{i_k}, Y_{i_j}\right\}, \qquad j=2,\ldots,n,
\end{equation}
so that the entries of $\theta$ corresponding to the untreated units are ``filled up'' in decreasing order of $Y$, i.e., $\theta_{i_j}=0$ unless $\theta_{i_k} = Y_{i_k}$ for $k=1,\ldots,j-1$.






\paragraph{Variant of \eqref{eq: conservative CI} used in \cite{miguel2004worms}} To estimate the number of infections that would occur if all of the schools were treated, we define $Y, X$, and $\theta$ as follows. Let $Y_i$ denote the number of infections observed in school $i$. Reversing the definition of $X$, let $X_i=0$ denotes that school $i$ receives the deworming treatment. Let $\theta$ denote the counterfactual outcomes that would occur if $X_i=0$ for all $i$. With $Y, X$, and $\theta$ thus defined, Assumption \ref{as: monotone}, which states that $\theta \leq Y$, means that treating all of the schools would not increase the infection counts over the observed values. A 95\% confidence upper bound on $\bar{\theta}$ can be found by solving \eqref{eq: conservative CI}.

To estimate the number of infections that would occur if none of the schools were treated, let $Y$ be defined as before; let $X_i=1$ denote that school $i$ receives deworming treatment; and let $\theta$ denote the counterfactual outcome that would occur if no schools receive treatment. In place of Assumption \ref{as: monotone}, we assume that $\theta_i \geq Y_i$, meaning that treating no schools would not reduce the infection counts below the observed values, and also that $\theta_i \leq S_i$, where $S_i$ is the total number of students at school $i$ that were measured in the 1999 survey. By similar reasoning as \eqref{eq: conservative CI}, in order to lower bound $\bar{\theta}$ we can solve 
\begin{align} \label{eq: reversed CI}
\min_{\theta \in \mathbb{Z}^N} &\quad  \hat{\theta} - t_\alpha\sqrt{\left(\frac{L}{N}\right) \frac{\hat{\sigma}^2}{N-L}} \\
\nonumber \st & \quad  Y_i \leq \theta_i \leq S_i \textrm{ for all } i,
\end{align}
where $\hat{\theta}$ and $\hat{\sigma}^2$ are defined as before. Similar to \eqref{eq: conservative CI}, the optimal $\theta$ must maximize $\hat{\sigma}^2$ along a level set of $\hat{\theta}$, so that 
\begin{align} \label{eq: reversed DP} 
\max_{\theta \in \mathbb{Z}^N} &\quad \sum_{i:X_i=0} \theta_i^2 \\
\nonumber \st & \quad \sum_{i:X_i=0} \theta_i = c \\
\nonumber & \quad   Y_i \leq \theta_i \leq S_i \textrm{ for all } i
\end{align}
can be solved for different values of $c$ to find the optimal $\theta$. 

The optimization problem \eqref{eq: reversed DP} can be formulated and solved as a dynamic programming problem. Generically, a simplified version of a dynamic program involves choosing a sequence of discrete decision variables $u_1,\ldots,u_T$, so as to control a sequence of state variables $s_0,\ldots,s_T$, where the initial state $s_0$ is given and $s_t = f_t(s_{t-1},u_t)$ for $t=1,\ldots,T$ and some set of functions $f_1,\ldots,f_T$ which model the state dynamics. A reward $g(u_t)$ is paid for each decision, and an final reward $G(s_T)$ is paid based on the final state. The goal is to choose $u_1,\ldots,u_T$ to maximize $G(s_T) + \sum_t g(u_t)$, thereby steering towards a high reward final state while also maintaining high rewards for each decision. A canonical algorithm to solve this problem is value iteration \cite{bertsekas1995dynamic}, which is also called backwards induction or Bellman's equation.

To formulate \eqref{eq: reversed DP} as a dynamic programming problem, let $T = n$ and let the decisions $u_1, \ldots, u_T$ equal $\theta_{i_1},\ldots, \theta_{i_n}$. Let $g(u_t) = u_t^2$, so that $\sum_t g(u_t)$ equals the objective of \eqref{eq: reversed DP}. Let $s_0=0$, and let $s_t = s_{t-1} + u_t$, so that $s_T = \sum_t u_t$, which equals $\sum_{i:X_i=0} \theta_i$. Let the final reward $G(s_T)$ equal $0$ if $s_T = c$, and $-\infty$ otherwise, thus enforcing the constraint that $\sum_{i:X_i=0} \theta_i = c$.

\section{Estimation Using $\Wb$} \label{appendix: wb}

\paragraph{Solution of \eqref{eq: exact} for $\Wb$} For $W = \Wb$, the $\alpha$-level critical value of $W$ is a function of $\sum_i \theta_i$, since $W$ is a $\operatorname{Hypergeometric}(\sum_i \theta_i, N - \sum_i \theta_i, L)$ random variable. Let $w_\alpha(\sum_i \theta_i)$ denote the $\alpha$-level critical value of $W$. It follows that \eqref{eq: exact} can be rewritten as
\begin{align} 
\nonumber \max_{\theta \in \{0,1\}^N} & \sum_{i=1}^N \theta_i \\
\st &\hskip.2cm \sum_{i:X_i=1} \theta_i \leq {w}_\alpha\left(\sum_{i=1}^N \theta_i\right) \label{eq: wb constraint 0} \\
 & \hskip.2cm \sum_{i:X_i=1} \theta_i \leq \sum_{i:X_i=1} Y_i \label{eq: wb constraint 1}\\
 & \hskip.2cm\sum_{i:X_i=0} \theta_i \leq \sum_{i:X_i=0} Y_i. \label{eq: wb constraint 2}, 
\end{align}
where \eqref{eq: wb constraint 1} and \eqref{eq: wb constraint 2} are consequences of $\theta \leq Y$. This optimization problem depends only the quantities $\sum_{i:X_i=1} \theta_i$ and $\sum_{i:X_i=0} \theta_i$. As these quantities are integer valued and bounded above and below, their optimal values can be easily found by exhaustive search.

\paragraph{Solution of \eqref{eq: weaker bound} for $\Wb$} For $W = \Wb$, the optimization problem \eqref{eq: weaker bound} can be rewritten as above, but with constraint \eqref{eq: wb constraint 1} removed. This removes the upper bound on $\sum_{i:X_i=1} \theta_i$. However, since $\sum_{i:X_i = 1}\theta_i \leq \sum_i X_i$, an upper bound still exists, so the optimal solution may be found by exhaustive search as before.

\section{Estimation Using $\Ws$} \label{appendix: ws}

For $W = \Ws$, the solution of of the optimization problem \eqref{eq: exact} is computationally hard. We present a conservative approximation of \eqref{eq: exact} that yields a larger confidence interval for $A$. The main steps of the approximation are to bound the critical value $w_\alpha(\theta)$ using a simpler expression, and to enclose the feasible region of \eqref{eq: exact} by linear inequalities. 

\paragraph{Preliminaries}

We will require the following basic identities. It can be seen that $\Ws(X;\theta)$ equals the average of $L$ samples drawn without replacement from the vector $K\theta$. Because the columns of $K$ sum to one, it holds that
\begin{equation} \label{eq: basic0}
\mathbb{E}\Ws(X;\theta) = \frac{1}{N}\sum_{i=1}^N \theta_i,
\end{equation}
where we note that the expectation $\mathbb{E} \equiv \mathbb{E}_X$ is taken over the random treatment $X$.

Let $u$ denote a unit sampled uniformly from $1,\ldots,N$. Let $1_u \in \{0,1\}^N$ denote the indicator function returning $1$ for unit $u$ and $0$ elsewhere. It follows that $\Ws(1_u;\theta)$ is equal in distribution to $\Ws(X;\theta)$ for $L=1$. For all $L$, it holds that 
\begin{align}
\label{eq: basic1}	\mathbb{E}\Ws(X;\theta) & = \mathbb{E}_u\Ws(1_u;\theta) \\
\label{eq: basic2}	\Var \Ws(X;\theta) & = \frac{N-L}{L(N-1)}\left( \mathbb{E}_u\left[\Ws(1_u;\theta)^2\right] - \left[\mathbb{E}_u\Ws(1_u;\theta)\right]^2\right),
\end{align}
where \eqref{eq: basic2} follows from basic properties of simple random sampling \cite[Eq. 2.5]{thompson2012sampling}.

\paragraph{Approximation of \eqref{eq: exact}}

By Chebychev's inequality, it holds for any choice of $W$ that
\begin{equation}\label{eq: chebychev}
	\mathbb{P}\left( \frac{W(X;\theta) - \mathbb{E}W(X;\theta)}{\left(\Var W(X;\theta)\right)^{1/2}} \geq \alpha^{-1/2}\right) \leq \alpha. 
\end{equation}
This is a highly conservative bound, but we use it here for simplicity and defer improvements for later discussion. 
Analogous to \eqref{eq: exact}, a one-sided $(1-\alpha)$ confidence interval for $\sum_i \theta_i$ is given by
\begin{align} \label{eq: clt}
\max_{\theta \in \{0,1\}^N} &\hskip.2cm  \frac{1}{N} \sum_{i=1}^N \theta_i \\
\nonumber \st &\hskip.2cm \frac{W(X;\theta) - \mathbb{E}W(X;\theta)}{\left(\Var W(X;\theta)\right)^{1/2}} \leq \alpha^{-1/2} \\
\nonumber & \hskip.2cm \theta_i \leq Y_i \textrm{ for all } i.
\end{align} 
To rewrite this problem with a smaller number of decision variables, let $m(y) \in \mathbb{R}^3$ denote the vector given by
\begin{equation*}
m_1(\theta) = \mathbb{E}_u W(1_u;\theta), \quad
m_2(\theta) = W(X;\theta), \quad \text{ and } \quad 
m_3(\theta) = \mathbb{E}\left[W(1_u;\theta)^2\right].
\end{equation*}	
Let $\mathcal{M} = \{m(\theta): \theta \leq Y\}$ denote the set of all achievable values for $m(\theta)$. Equating terms and using \eqref{eq: basic0}-\eqref{eq: basic2}, the optimization problem \eqref{eq: clt} can be restated as 
\begin{align} 
\max_{m \in \mathbb{R}^3} & \hskip.2cm m_1 \label{eq: upper M}\\
\nonumber \st &\hskip.2cm \frac{m_2 - m_1}{\left(m_3 - m_1^2\right)^{1/2}} \leq \left(\alpha L\frac{N-1}{N-L}\right)^{-1/2} \\
\nonumber &\hskip.2cm m \in \mathcal{M}.
\end{align}
While this optimization problem has only 3 decision variables, it is hard to optimize because the constraint $m \in \mathcal{M}$ is difficult to check. As a relaxation, we will replace the constraint $m \in \mathcal{M}$ by a weaker constraint $m \in \mathcal{P}$, where $\mathcal{P}$ is a polyhedron that contains $\mathcal{M}$, and which can be represented by a tractable number of linear inequalities. Let $f^*(\lambda)$ denote the maximum inner product between $\lambda \in \mathbb{R}^3$ and $m(\theta) \in \mathcal{M}$:
\begin{equation*} \label{eq: fstar}
f^*(\lambda) = \max_{\theta\in \{0,1\}^N} \lambda^Tm(\theta) \quad \st \ \theta \leq Y.
\end{equation*}
Given a set $\Lambda \subset \mathbb{R}^3$, let $\mathcal{P}_\Lambda$ denote the set $\{m: \lambda^Tm \leq f^*(\lambda)\ \textrm{ for all } \lambda \in \Lambda\}$. Since $\lambda^T m \leq f^*(\lambda)$ for all $m \in \mathcal{M}$, it follows that $\mathcal{P}_\Lambda$ contains $\mathcal{M}$. Hence the following optimization problem upper bounds \eqref{eq: upper M}, yielding a conservative confidence interval:
\begin{align} 
\label{eq: upper lagrange} \max_{m \in \mathbb{R}^3} & \hskip.2cm m_1 \\
\nonumber \st &\hskip.2cm \frac{m_2 - m_1}{\left(m_3 - m_1^2\right)^{1/2}} \leq \left(\alpha L\frac{N-1}{N-L}\right)^{-1/2} \label{eq: upper lagrange 1}\\
\nonumber &\hskip.2cm \lambda^Tm \leq f^*(\lambda),\ \forall\ \lambda \in \Lambda.
\end{align}
This optimization problem is low dimensional. As a result, it can be practically solved by a grid-based search over the feasible region, provided that $f^*(\lambda)$ is known for all $\lambda \in \Lambda$.

\paragraph{Computation of $f^*(\lambda)$} To solve \eqref{eq: upper lagrange}, we must compute $f^*(\lambda)$ for all $\lambda \in \Lambda$. For $W = \Ws$, it holds by the following identities,
\begin{equation*}
\mathbb{E}_u\Ws(1_u;\theta) = \frac{\obf^T K \theta}{N}, \qquad 
\mathbb{E}_u\left[ \Ws(1_u;\theta)^2\right] = \frac{\theta^T K^T K \theta}{N}, \quad \textrm{and} \quad
W(X;\theta) = \frac{X^T K \theta}{L},
\end{equation*}
that we may write $f^*(\lambda)$ as 
\begin{align} \label{eq: fstar expanded}
f^*(\lambda) & = \max_{\theta \in \{0,1\}^N} \lambda_1 \frac{\obf^T K \theta}{N} + \lambda_2 \frac{X^TK \theta}{L} + \lambda_3 \frac{\theta^T K^T K \theta}{N}, \\
\nonumber & \hskip.3cm \st\ \theta_i \leq Y_i \textrm{ for all } i.
\end{align}
For nonnegative $K$ and $\lambda_3$, \eqref{eq: fstar expanded} can be transformed into a canonical optimization problem of finding an ``$s$-$t$ min cut'' in a graph. The transformation, described in Appendix \ref{appendix: min cut}, was originally proposed in \cite{greig1989exact} for image denoising. After the transformation, the min cut problem can be solved by linear programming or the Ford-Fulkerson algorithm, which runs in $O(n^3)$ time where $n = \sum_i Y_i$. \cite{papadimitriou1998combinatorial} 

\begin{figure} 
 \centering
 \begin{subfigure}[t]{.3\textwidth}
	\includegraphics[width=1.95in]{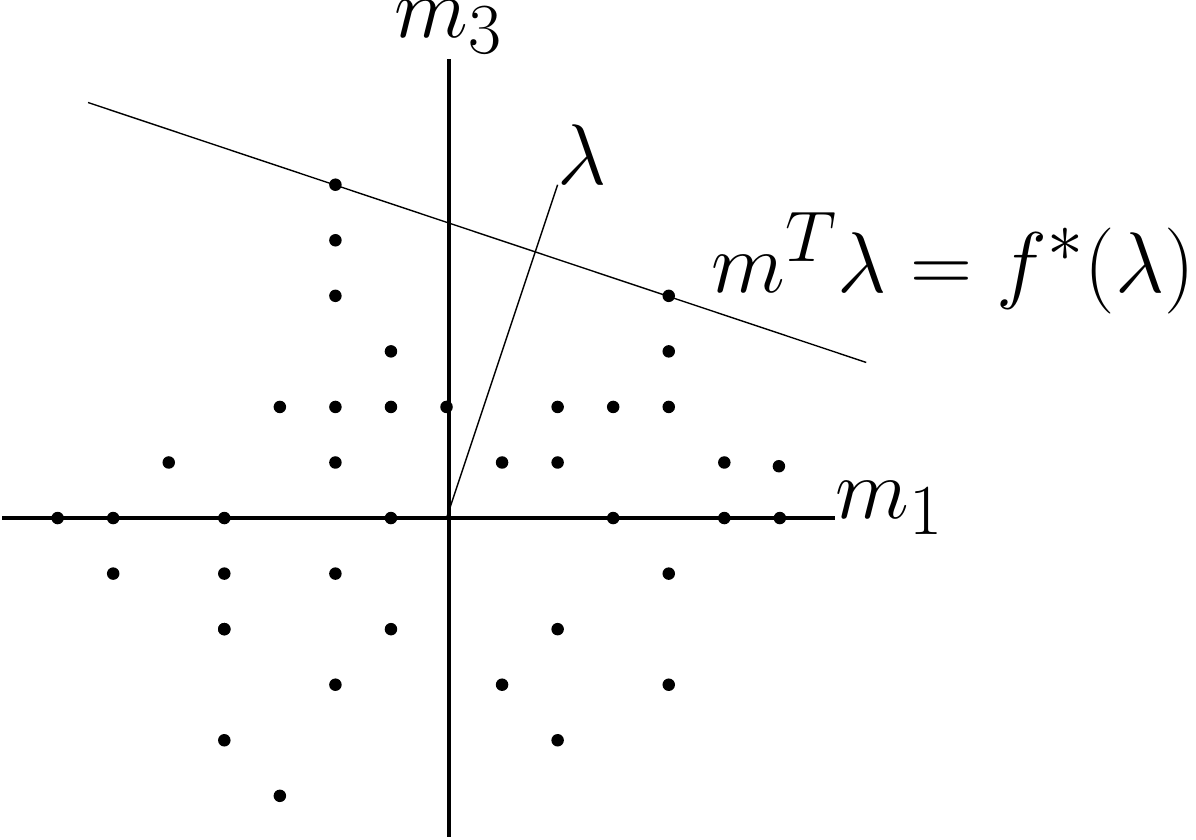}
	\caption{$\mathcal{M}$ and $m^T\lambda = f^*(\lambda)$}
\label{fig: lambda 1}
 \end{subfigure}
 \begin{subfigure}[t]{.3\textwidth}
	\includegraphics[width=1.6in]{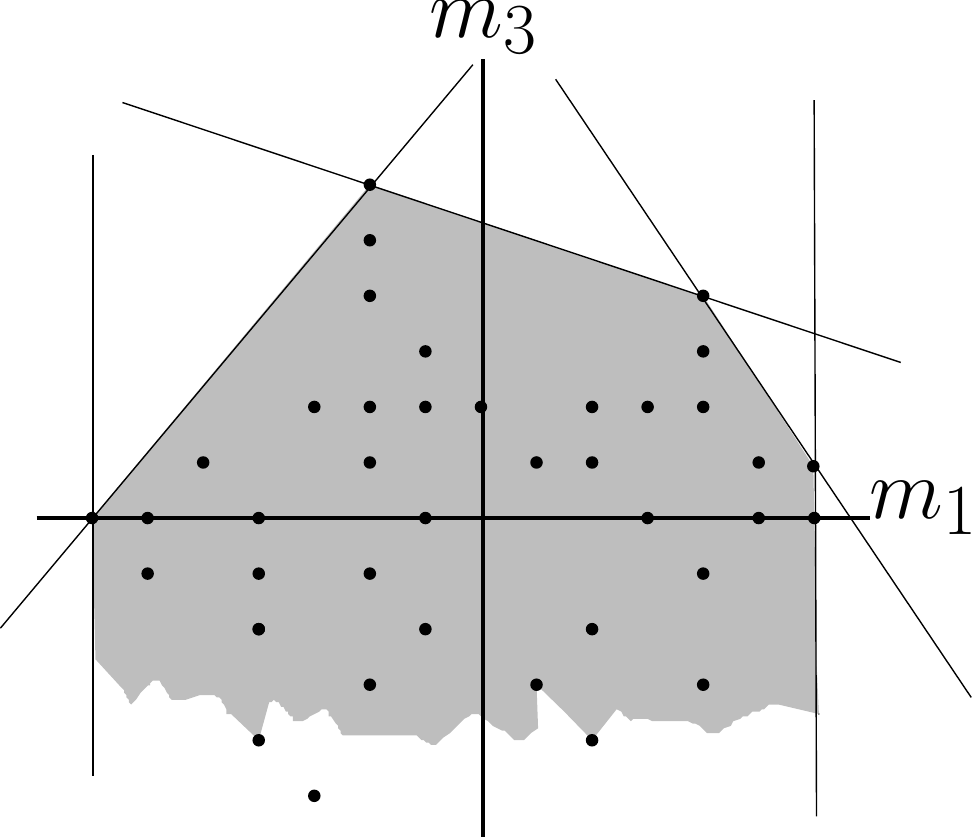}
	\caption{$\mathcal{P}_\Lambda$}
\label{fig: lambda 2}
 \end{subfigure}
 \begin{subfigure}[t]{.3\textwidth}
	\includegraphics[width=1.9in]{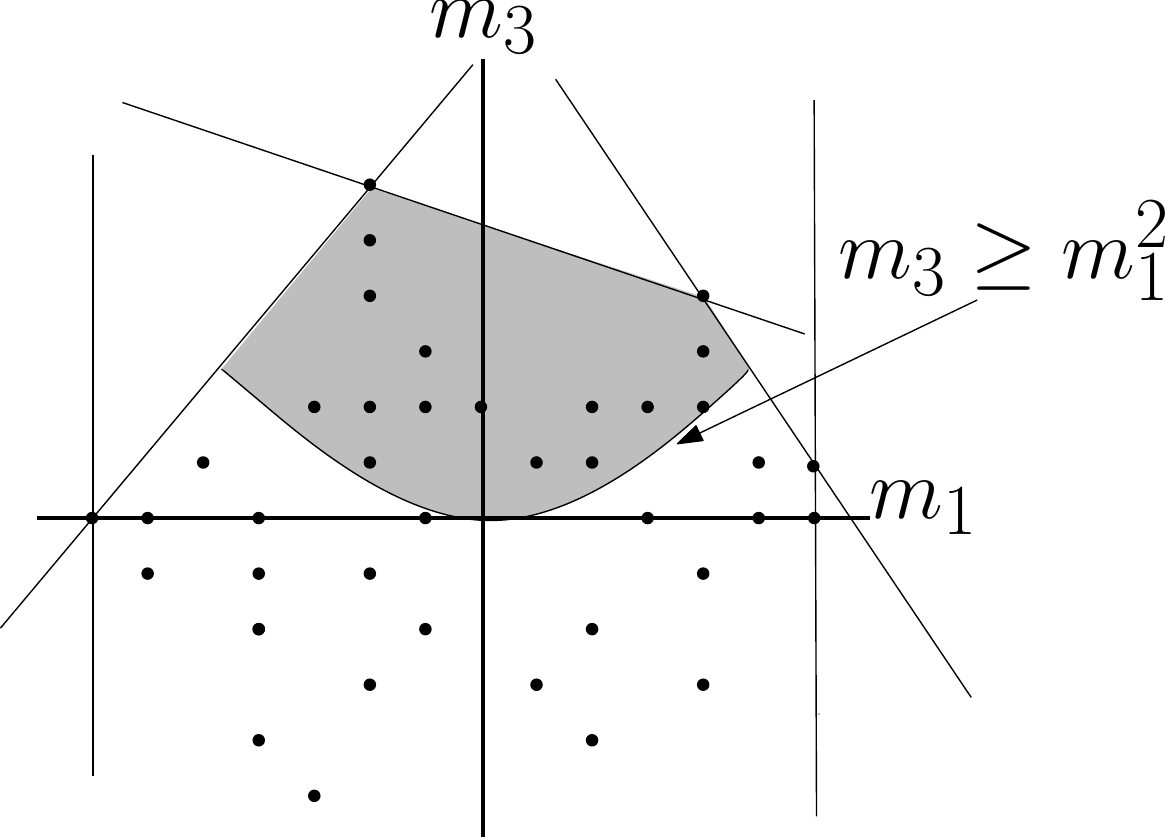}
	\caption{$\mathcal{P}_\Lambda \cap \{m_3 \geq m_1^2\}$}
\label{fig: lambda 3}
 \end{subfigure}
 \caption{Cartoon depiction of \eqref{eq: upper lagrange}, showing dimensions $m_1$ and $m_3$ only. (a) shows $\mathcal{M}$ (as dots), and a supporting hyperplane in a direction $\lambda$. (b) shows $\mathcal{P}_\Lambda$ (as shaded region), which may equal the convex hull of $\mathcal{M}$ in all directions $\lambda$ satisfying $\lambda_3 \geq 0$. (c) shows the intersection of $\mathcal{P}_\Lambda$ and the constraint $m_3 \geq m_1^2$. This constraint is implicit in \eqref{eq: upper lagrange}, since otherwise $(m_3 - m_1^2)^{-1/2}$ would not be real-valued. \label{fig: lambda}}
\end{figure}

\paragraph{Selection of $\Lambda$}
Figure \ref{fig: lambda} gives a geometric picture of the role of $\mathcal{M}$ and $\mathcal{P}_\Lambda$ in determining the feasible region of \eqref{eq: upper lagrange}. The set $\Lambda$ must satisfy $\lambda_3 \geq 0$ for all $\lambda \in \Lambda$, since $f^*(\lambda)$ cannot be efficiently computed otherwise. By definition, each half-space $H_\lambda = \{m: \lambda^T m \leq f^*(\lambda)\}$ equals a supporting hyperplane of the set $\mathcal{M}$ in the direction $\frac{\lambda}{\|\lambda\|}$. This implies that $H_\lambda = H_{c \lambda}$ when $c$ is a positive scalar. As a result, a reasonable strategy is to choose $\Lambda$ to cover the allowable directions $\{\lambda: \|\lambda\|=1, \lambda_3 \geq 0\}$ as densely as possible, so that $\mathcal{P}_\Lambda$ approximates the convex hull of $\mathcal{M}$ in those directions.


\paragraph{Reducing conservativeness} Chebychev's inequality gives a very conservative approximation to the critical value of the test statistic. Because $\Ws(X;\theta)$ is a sample average, a normal approximation may yield a better estimate of its critical value. That is, it may hold that 
\begin{equation*}
	\mathbb{P}\left( \frac{\Ws(X;\theta) - \mathbb{E}\Ws(X;\theta)}{\left(\Var \Ws(X;\theta)\right)^{1/2}} \geq z_\alpha \right) \approx \alpha, 
\end{equation*}
where $z_\alpha$ is the upper critical value of a standard normal. Using this approximation leads to the following optimization problem 
\begin{align} 
\label{eq: upper lagrange CLT} \max_{m \in \mathbb{R}^3} & \hskip.2cm m_1 \\
\nonumber \st &\hskip.2cm \frac{m_2 - m_1}{\left(m_3 - m_1^2\right)^{1/2}} \leq  z_\alpha L^{-1/2}\\
\nonumber &\hskip.2cm \lambda^Tm \leq f^*(\lambda),\ \forall\ \lambda \in \Lambda.
\end{align}

\paragraph{Summary of method} Given binary observations $Y$, treatment assignment $X$, and network information $G$, the method entails the following steps:
\begin{enumerate}
\item Choose a smoothing matrix $K$, for example by choosing values of $d_{\max, K}$ and $\sigma_K$.
\item Choose a set $\Lambda \subset \mathbb{R}^3$ such that $\lambda_3 \geq 0$ for all $\lambda \in \Lambda$. This will ultimately induce the set $\mathcal{P}$ which relaxes the actual feasible region.
\item For each $\lambda \in \Lambda$, compute $f^*(\lambda)$ by solving \eqref{eq: fstar expanded}. The solution of \eqref{eq: fstar expanded} is discussed in Appendix \ref{appendix: min cut}. 
\item Solve \eqref{eq: upper lagrange} or \eqref{eq: upper lagrange CLT} to the desired level of precision. This is done by discretizing the feasible region of \eqref{eq: upper lagrange} or \eqref{eq: upper lagrange CLT} along a grid, and checking every grid point. Because the objective is linear and the feasible region is 3-dimensional, the number of grid points that must be checked increases cubically with the desired precision.  The best solution is an upper bound on $\sum_i \theta_i$, up to the precision of the grid search.
\end{enumerate}

\section{Transformation of $f^*(\lambda)$ to min-cut problem} \label{appendix: min cut}

Given a nonnegative matrix $A \in \mathbb{R}^{d \times d}$ with zero diagonal, and $s,t \in 1,\ldots,d$, the s-t min cut problem is 
\begin{align} 
\label{eq: min cut} \min_{x \in \{0,1\}^d} & \sum_{i\neq j} A_{ij} x_i(1-x_j) \\
\nonumber \st &\hskip.2cm x_s=1, x_t=0.
\end{align}
The interpretation of \eqref{eq: min cut} is that $A$ denotes a weighted adjacency matrix of a network, and $x$ divides the nodes $1,\ldots,d$ into two groups, with $s$ and $t$ in separate groups, so as to minimize the sum of the weighted edges that are ``cut'' by the division. This problem is polynomially solvable by the Ford-Fulkerson algorithm and also by linear programming \cite{papadimitriou1998combinatorial}.

To transform $f^*(\lambda)$ into the form of \eqref{eq: min cut}, we observe that 
\begin{align*} 
f^*(\lambda) & = \max_{\theta \in \{0,1\}^N} \lambda_1 \frac{\obf^T K \theta}{N} + \lambda_2 \frac{X^TK \theta}{L} + \lambda_3 \frac{\theta^T K^T K \theta}{N}, \\
& \hskip.3cm \st\ \theta_i \leq Y_i \textrm{ for all }i,
\end{align*}
may be rewritten as
\[ \max_{x \in \{0,1\}^d} x^TMx + b^T x + c,\]
for some $d>0$, $b \in \mathbb{R}^d$, $c \in \mathbb{R}$, and nonnegative matrix $M$,  where the decision variable $x$ corresponds to the free elements in $y$, i.e., those in $\{i: Y_i=1\}$ . Following \cite{greig1989exact}, we transform this to a min-cut problem by observing that
\begin{align}
\nonumber x^T M x + b^T x& = -\sum_{i,j} \left( M_{ij} x_i(1-x_j) - M_{ij} x_{i} \right) + \sum_i b_i x_i \\
 & = -\sum_{i\neq j} M_{ij} x_i(1-x_j) + \sum_i x_i\left(b_i+\sum_j M_{ij}\right). \label{eq: min cut step1}
	\end{align}
Let $\gamma_i = b_i + \sum_j M_{ij}$. Then maximizing \eqref{eq: min cut step1} is equivalent to 
\begin{align} \label{eq: min cut step2}
	\max_{x\in \{0,1\}^d} -\sum_{i\neq j} M_{ij} x_i(1-x_j) - \sum_{i:\gamma_i \geq 0} |\gamma_i|(1-x_i) + \sum_{i:\gamma_i <0} |\gamma_i|x_i. 
\end{align}
Let $s=d+1$, $t=d+2$, and let $x_s=1,x_t=0$.	We can rewrite \eqref{eq: min cut step2} as
\[	\max_{x\in \{0,1\}^d} 	-\sum_{i\neq j} M_{ij} x_i(1-x_j) - \sum_{i:\gamma_i \geq 0} |\gamma_i|(1-x_i)x_s + \sum_{i:\gamma_i <0} |\gamma_i|x_i(1-x_t),\]
which can be rewritten as \eqref{eq: min cut} for some nonnegative $A \in \mathbb{R}^{d+2 \times d+2}$ with zero diagonal.



\bibliographystyle{apalike}
\bibliography{bibfile}
\end{document}